\documentclass{emulateapj}
\usepackage{apjfonts}
\usepackage{graphicx}

\def\apj {ApJ}

\def\apjs {ApJS}
\def\aj {AJ}
\def\mnras {MNRAS}
\def\aap {A\&A}
\def\nat {Nat}
\def\sci {Sci}
\def\araa {ARAA}

\def\apss {A\&SS}
\def\bea{\begin{eqnarray}}
\def\eea{\end{eqnarray}}
\def\bee{\begin{equation}}
\def\eee{\end{equation}}
\def\bef{\begin{figure}}
\def\eef{\end{figure}}
\def\befs{\begin{figure*}}
\def\eefs{\end{figure*}}
\def\be{\begin{equation}}
\def\ee{\end{equation}}

\def\NII  {[${\rm N\:\scriptstyle II}$]}
\def\SII  {[${\rm S\:\scriptstyle II}$]}

\def\OI {[${\rm O\:\scriptstyle I}$]}

\def\OIII {[${\rm O\:\scriptstyle III}$]}
\def\HI   {${\rm H\:\scriptstyle\rm I}$}

\def\HeII   {${\rm He\:\scriptstyle\rm II}$}

\def\ff   {\ifmmode{f}\else{$f$}\fi}

\def\rmax {\ifmmode{r_{\rm max}}\else{$r_{\rm max}$}\fi}
\def\zmax {\ifmmode{z_{\rm max}}\else{$z_{\rm max}$}\fi}

\def\fCNM {\ifmmode{f^{\rm CNM}}\else{$f^{\rm CNM}$}\fi}
\def\fCMM {\ifmmode{f^{\rm CMM}}\else{$f^{\rm CMM}$}\fi}
\def\fWNM {\ifmmode{f^{\rm WNM}}\else{$f^{\rm WNM}$}\fi}
\def\fWIM {\ifmmode{f^{\rm WIM}}\else{$f^{\rm WIM}$}\fi}
\def\fCM  {\ifmmode{f^{\rm CM}} \else{$f^{\rm CM}$}\fi}
\def\fWM  {\ifmmode{f^{\rm WM}} \else{$f^{\rm WM}$}\fi}

\def\noCNM {\ifmmode{n_o^{\rm CNM}}\else{$n_o^{\rm CNM}$}\fi}
\def\noCMM {\ifmmode{n_o^{\rm CMM}}\else{$n_o^{\rm CMM}$}\fi}
\def\noWNM {\ifmmode{n_o^{\rm WNM}}\else{$n_o^{\rm WNM}$}\fi}
\def\noWIM {\ifmmode{n_o^{\rm WIM}}\else{$n_o^{\rm WIM}$}\fi}
\def\noCM  {\ifmmode{n_o^{\rm CM}} \else{$n_o^{\rm CM}$}\fi}
\def\noWM  {\ifmmode{n_o^{\rm WM}} \else{$n_o^{\rm WM}$}\fi}

\def\nnCNM {\ifmmode{n^{\rm CNM}}\else{$n^{\rm CNM}$}\fi}
\def\nnCMM {\ifmmode{n^{\rm CMM}}\else{$n^{\rm CMM}$}\fi}
\def\nnWNM {\ifmmode{n^{\rm WNM}}\else{$n^{\rm WNM}$}\fi}
\def\nnWIM {\ifmmode{n^{\rm WIM}}\else{$n^{\rm WIM}$}\fi}
\def\nnCM  {\ifmmode{n^{\rm CM}} \else{$n^{\rm CM}$}\fi}
\def\nnWM  {\ifmmode{n^{\rm WM}} \else{$n^{\rm WM}$}\fi}

\def\zCNM {\ifmmode{z_o^{\rm CNM}}\else{$z_o^{\rm CNM}$}\fi}
\def\zCMM {\ifmmode{z_o^{\rm CMM}}\else{$z_o^{\rm CMM}$}\fi}
\def\zWNM {\ifmmode{z_o^{\rm WNM}}\else{$z_o^{\rm WNM}$}\fi}
\def\zWIM {\ifmmode{z_o^{\rm WIM}}\else{$z_o^{\rm WIM}$}\fi}
\def\zCM  {\ifmmode{z_o^{\rm CM}} \else{$z_o^{\rm CM}$}\fi}
\def\zWM  {\ifmmode{z_o^{\rm WM}} \else{$z_o^{\rm WM}$}\fi}

\def\nH {\ifmmode{\langle n_H \rangle}\else{$\langle n_H \rangle$}\fi}
\def\Em {\ifmmode{E_m}\else{$E_m$}\fi}
\def\deg   {\ifmmode{^\circ}\else{$^\circ$}\fi}
\def\HH    {\ifmmode{\cal H}\else{${\cal H}$}\fi}
\def\dH    {\ifmmode{\delta\cal H}\else{$\delta{\cal H}$}\fi}
\def\Hdisk {\ifmmode{\cal H}_{\rm disk}\else{${\cal H}_{\rm disk}$}\fi}
\def\Hdiskp {\ifmmode{\cal H}^{'}_{\rm disk}\else{${\cal H}^{'}_{\rm disk}$}\fi}
\def\Hsky  {\ifmmode{\cal H}_{\rm sky}\else{${\cal H}_{\rm sky}$}\fi}

\def\gta{\;\lower 0.5ex\hbox{$\buildrel > \over \sim\ $}}
\def\lta{\;\lower 0.5ex\hbox{$\buildrel < \over \sim\ $}}          
\def \NH {\ifmmode{\rm N}_{\scriptscriptstyle H}\else{N$_{\scriptscriptstyle H}$}\fi}

\def\deg{\hbox{${}^\circ$}}

\def\las{\mathrel{\hbox{\rlap{\hbox{\lower4pt
        \hbox{$\sim$}}}\hbox{$<$}}}}
\def\gas{\mathrel{\hbox{\rlap{\hbox{\lower4pt
            \hbox{$\sim$}}}\hbox{$>$}}}}

\begin{document}

\title{Fossil imprint of a powerful flare at the Galactic Centre along
  the Magellanic Stream}  

\author{J. Bland-Hawthorn\altaffilmark{1}}
\affil{Sydney Institute for Astronomy, School of Physics A28,
  University of Sydney, NSW 2006, Australia} 
  \footnote{Visiting Fellow, Australian Astronomical Observatory, 105 Delhi Rd, North Ryde, NSW 2113, Australia}
\author{Philip R. Maloney} 
\affil{CASA, University of Colorado, Boulder, CO 80309-0389, USA}
\author{Ralph S. Sutherland}
\affil{Mount Stromlo Observatory, Australia National University,
  Woden, ACT 2611, Australia}
\author{G. J. Madsen}
\affil{Institute of Astronomy, University of Cambridge, Madingley Rd,
  Cambridge, CB3 0HA, UK} 

\begin{abstract}
\vspace{0.1in} The {\it Fermi} satellite discovery of the gamma-ray
emitting bubbles extending 50$^{\circ}$ (10 kpc) from the Galactic
Centre has revitalized earlier claims that our Galaxy has undergone an
explosive episode in the recent past.  We now explore a new constraint
on such activity. The Magellanic Stream is a clumpy gaseous structure
free of stars trailing behind the Magellanic Clouds, passing over the
South Galactic Pole (SGP) at a distance of at least 50$-$100 kpc from the
Galactic Centre. Several groups have detected faint H$\alpha$ emission
along the Magellanic Stream ($1.1\pm 0.3\times 10^{-18}$ erg cm$^{-2}$
s$^{-1}$ arcsec$^{-2}$) that is a factor of 5 too bright to have been
produced by the Galactic stellar population. The brightest emission is
confined to a cone with half angle $\theta_{1/2} \approx 25^\circ$
roughly centred on the SGP.  Time-dependent models of Stream clouds
exposed to a flare in ionising photon flux show that the ionised gas
must recombine and cool for a time interval $T_o = 0.6 - 2.9$ Myr for
the emitted H$\alpha$ surface brightness to drop to the observed
level. A nuclear starburst is ruled out by the low star formation
rates across the inner Galaxy, and the non-existence of starburst
ionisation cones in external galaxies extending more than a few
kiloparsecs.  Sgr A$^\star$ is a more likely candidate because it is
two orders of magnitude more efficient at converting gas to UV
radiation. The central black hole ($M_\bullet \approx 4\times10^6$
M$_\odot$) can supply the required ionising luminosity with a fraction
of the Eddington accretion rate
($f_E \sim 0.03-0.3$, depending on uncertain factors, e.g., Stream
distance), typical of Seyfert
galaxies. In support of nuclear activity, the H$\alpha$ emission along
the Stream has a polar angle dependence peaking close to the
SGP. Moreover, it is now generally accepted that the Stream over the
SGP must be further than the Magellanic Clouds. At the lower halo gas
densities, shocks become too ineffective and are unlikely to give rise
to a polar angle dependence in the H$\alpha$ emission.  Thus it is
plausible that the Stream H$\alpha$ emission arose from a `Seyfert flare' that
was active $1-3$ Myr ago, consistent with the cosmic ray lifetime in
the {\it Fermi} bubbles. Such a flare may have been causally linked to one 
of the episodes of massive star formation triggered in close proximity to 
Sgr A$^\star$ within the last few Myr. Sgr A$^\star$ activity today is greatly
suppressed (70-80 dB) relative to the Seyfert outburst. The rapid
change over a huge dynamic range in ionising luminosity
argues for a compact UV source with an extremely efficient (presumably
magneto-hydrodynamic) `drip line' onto the accretion disk.
\end{abstract}

\section{Introduction}
Nuclear activity powered by a supermassive black hole
is a remarkable phenomenon that allows galaxies to be
observed to at least a redshift $z\approx 7$ (e.g. Mortlock et al
2011). Evidence is beginning to emerge that our Galaxy has experienced
possibly related episodes in the recent past. The proximity of the
Galactic Centre provides us with an opportunity to study this
activity in unprecedented detail.

The first evidence for a large-scale bipolar outflow came from
extended bipolar {\it ROSAT} 1.5 keV X-ray and {\it MSX} 8.3$\mu$m
mid-infrared emission observed to be associated with the Galactic
Centre (Bland-Hawthorn \& Cohen 2003). Two observations made clear
that this activity {\it must} be associated with the centre of the
Galaxy: (i) the bipolar structure is not visible in the diffuse {\it
ROSAT} 0.2$-$0.5 keV data because the disk is optically thick at these
energies; (ii) a hard X-ray bipolar counterpart has never been
observed from a blow-out due to a young star cluster, thereby ruling
out any association with a spiral arm along the line of sight. Further
support for the large-scale wind picture comes from a population of
entrained \HI\ clouds (McClure-Griffiths et al 2013) and from the kinematics of
low-column halo clouds observed in absorption along quasar sight lines
(Keeney et al 2006). In summary, the wind energetics are estimated to
be roughly 10$^{55}$ erg visible over 20$^\circ$ (5 kpc in radius).

In 2010, spectacular evidence for a powerful nuclear event came from
{\it Fermi} gamma-ray satellite observations (1$-$100 GeV) of giant
bipolar bubbles extending 50$^\circ$ (10 kpc) from the Galactic Centre
(Su et al 2010). The source of the bubbles, whether related to
starburst or AGN activity, remains hotly contested (Zubovas et al
2011; Su \& Finkbeiner 2012; Carretti et al 2013). The bubbles appear
to be associated with the very extended radio emission (`haze') first
identified in {\it WMAP} microwave observations (Finkbeiner 2004) and
are clearly associated with the bipolar X-ray structures
(Bland-Hawthorn \& Cohen 2003). One interpretation is that the
gamma-ray photons arise through inverse Compton scattering of the
interstellar and cosmic background radiation field by high energy
cosmic rays (10$-$100 GeV) from the black hole accretion disk (Su et
al 2010; Dobler et al 2010). In this scenario, the cosmic ray cooling
times $T_{CR}$ are of order a few million years, suggesting powerful
nuclear activity on a similar timescale. This picture implicates very
fast nuclear winds with speeds of order $\sim 10$ kpc/$T_{CR}$ $\sim$
10$^4$ km s$^{-1}$.

Guo \& Mathews (2012) have recently challenged the wind picture on the
grounds that the diffusion coefficient for cosmic rays advected in
winds is much lower than required to explain the Fermi
bubbles. Instead, they suggest the cosmic rays are carried by bipolar
jets from an active galactic nucleus (AGN) which inflated the Fermi
bubbles. Spectacular examples of this phenomenon do exist in nearby
Seyfert galaxies (e.g. 0313-192; Keel et al 2006).  Preliminary
evidence for nuclear jets at the Galactic Centre has been discussed by
several authors (Su \& Finkbeiner 2012; Yusef-Zadeh et al 2012).  In
the Guo \& Mathews model, the jets formed 1$-$3 Myr ago and endured
for $0.1-0.5$ Myr with a total energy in the range 10$^{55-57}$ erg.

The supermassive black hole associated with the Galactic Centre source
Sgr A$^\star$ has a well established mass\footnote{For a review of all
estimates of $M_\bullet$ to date, see Kormendy \& Ho (2013).} with
10\% uncertainty ($M_\bullet \approx 4\times 10^6$ M$_\odot$; Genzel
et al 2003; Meyer et al 2012).  At the present time, little is known
about past nuclear activity. It is likely that the black hole was far
more active before a redshift of unity when galactic accretion was at
its peak. But even at the present epoch, there is good evidence for
enhanced nuclear activity in interacting $L^\star$ galaxies (q.v. Wild
et al 2013; Rupke \& Veilleux 2013).  Direct evidence that the nuclear
regions were much brighter in the past comes from {\it ASCA} 2$-$10
keV observations of circumnuclear clouds (Sunyaev et al 1993; Koyama
et al 1996) with indications that Sgr A$^\star$ was 10$^5$ times more
active within the past 10$^3$ years (q.v. Ponti et al 2010). More
compelling evidence on much longer timescales comes from the existence
of the Fermi bubbles.

We now show that if our Galaxy went through a Seyfert phase in the
recent past, it could conceivably have been so UV bright that it lit
up the Magellanic Stream over the South Galactic Pole (SGP) through
photoionisation.  Interestingly, the Magellanic Stream has detectable
H$\alpha$ emission along its length that is at least 5 times more
luminous than can be explained by UV escaping from the Galaxy (see \S
2). Ionisation cones have been observed in several dozen Seyfert
galaxies to date.  Arguably, the most spectacular example is the S0
galaxy NGC 5252 (Tadhunter \& Tsvetanov 1989): orbiting gas streams up
to 30 kpc in radius are lit up along bipolar cones due to
the nuclear UV flux (Tsvetanov et al 1996). Kreimeyer \& Veilleux
(2013) have discovered ionisation cones in MR2251-178, a nearby
quasar with a weak double-lobed radio source; non-thermal photoionisation
is seen out to 90~kpc in radius emphasising the extraordinary
reach of AGN activity to the present day.

In principle, the Stream H$\alpha$ emission could have been produced
by a starburst event in the Galactic Centre, rather than by a Seyfert
flare.  However, as we discuss in Appendix B, the required star
formation rate of such a starburst is at least two orders of magnitude
larger than allowed by the star formation history of the Galactic
Centre. An accretion flare from Sgr A$^\star$ is a much more probable
candidate for the ionisation source because (a) an accretion disk
converts gas to ionising radiation with much greater efficiency than
star formation, thus minimizing the fuelling requirements; (b) there
is an abundance of material in the vicinity of Sgr A* to fuel such an
outburst, and (c) a rapid decline in the ionising luminosity, needed
in both starburst and AGN models to reconcile the present-day lack of
activity with the magnitude of the required flare, is prohibitively
difficult for starburst models but achievable (and interestingly
challenging) for accretion disk models (cf. \S 5.1). Regardless of the
true origin of the Stream's H$\alpha$ emission, we show that its
brightness is a powerful constraint on recent nuclear activity.

In \S 2, we describe basic properties of the Magellanic Stream and
derive the levels of ionisation required to explain the
observations. In \S 3, we carry out time dependent ionisation
calculations and relate to past AGN activity. We suggest follow-up
observations in \S 4 and discuss the implications of our findings in
\S 5. We conclude the paper with extended supplementary material on the gas
physics, ionisation requirements and the ionisation spectrum in three
appendices.

\begin{figure*}[htbp]
   \centering
   \includegraphics[scale=0.4,trim=0mm 0mm 0mm 0mm]{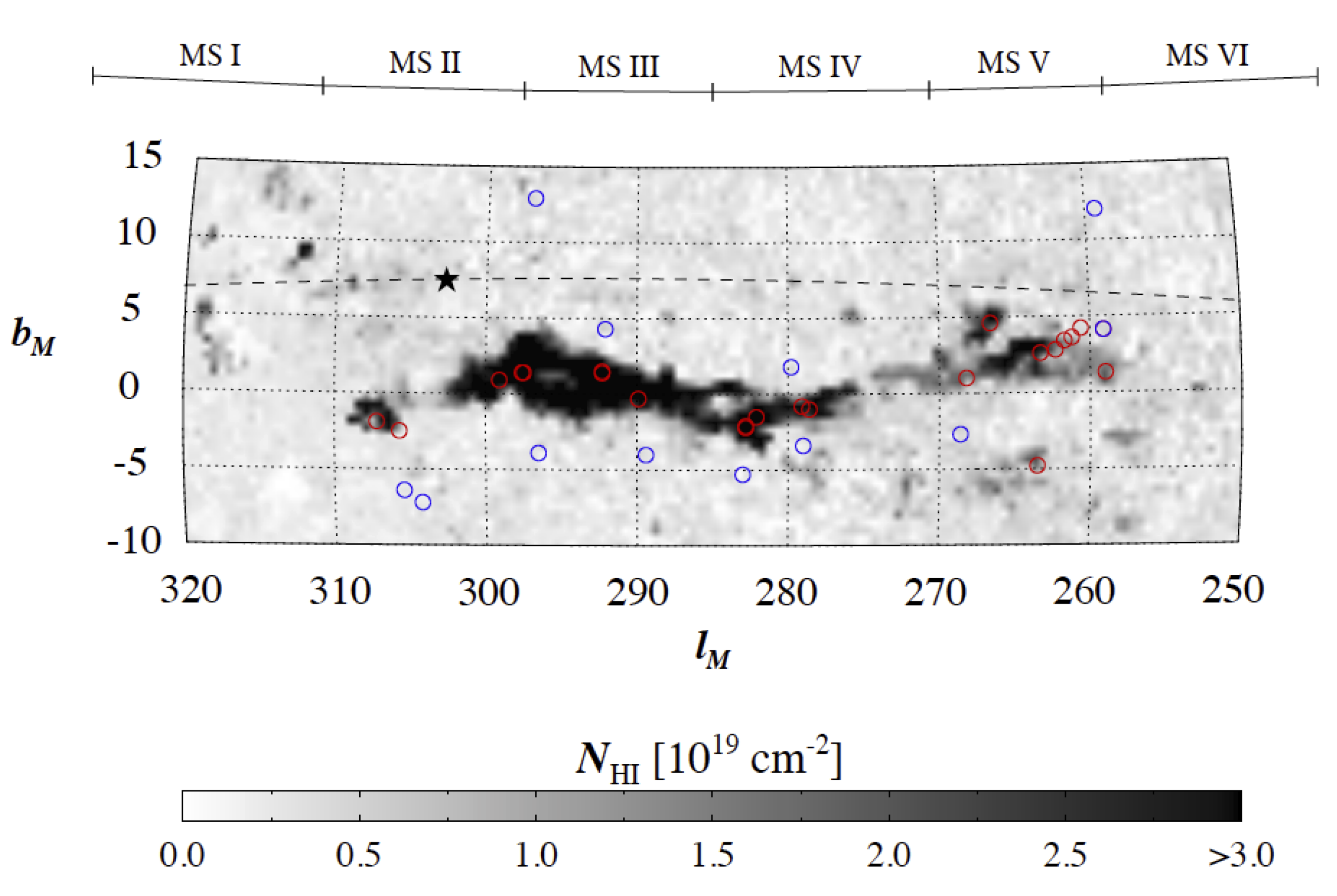}
   \includegraphics[scale=0.45,trim=0mm 0mm 0mm 0mm]{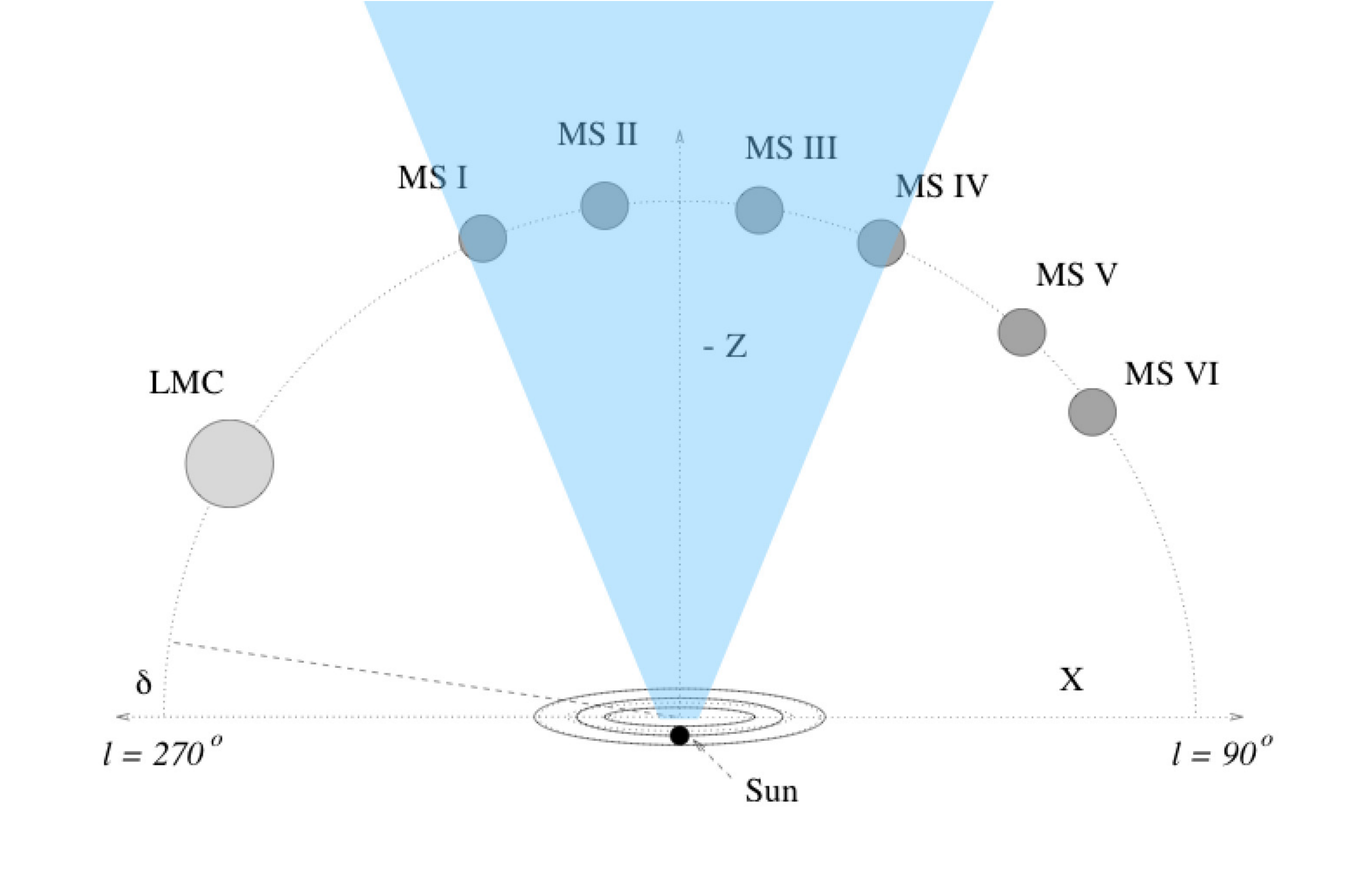}
      \caption{ {\it (Top)} HI column density map of the Magellanic
   Stream (adapted from Madsen 2012).  The coordinates are in
   Magellanic longitude and latitude ($\ell_M, b_M$), as defined by
   Nidever et al (2008), and the asterisk indicates the SGP. The
   linear greyscale shows the column density of HI from the LAB survey
   (Kalberla et al. 2005), with the 21cm emission integrated over the
   velocity range of -450 km s$^{-1}$ $<$ $v_{LSR}$ $<$ -100 km
   s$^{-1}$. The location and 1$^\circ$ field of view of the new WHAM
   observations are shown as red and blue circles, corresponding to
   target and sky observations, respectively. The approximate
   longitudinal extents of the six Stream complexes (identified by
   Mathewson \& Ford 1984) are shown on the top of the figure.  {\it
   (Bottom)} An illustration of the LMC and the dominant clouds in the
   Magellanic Stream (Mathewson \& Ford 1984) projected onto the
   Galactic $X$-$Z$ plane. The orbit of the Stream lies close to the
   great circle whose Galactic longitude is $\ell = 280^\circ$ (lhs of
   SGP) and $\ell = 100^\circ$ (rhs of SGP) shown as a dashed
   line. The blue fan illustrates the proposed ionisation cone from
   the Galactic Centre.}
   \label{f:MS}
   \medskip
\end{figure*}

\section{Experiment}

\noindent{\sl Target.}  
The Magellanic Stream (Fig.~\ref{f:MS}$a$) lies along a great arc that
extends for more than 150$^\circ$ (Mathewson, Cleary \& Murray 1974;
Putman et al 1998; Nidever et al 2008).  Fig.~\ref{f:MS}$b$
illustrates the relationship of the LMC to the Magellanic Stream above
the Galactic disk along a circular orbit originating from the Lagrangian
point between the LMC and the SMC, at a Galactocentric distance of 55
kpc (Mathewson \& Ford 1984). More recent simulations tend to suggest
that the LMC-SMC system is infalling for the first time with an orbital
period of order a Hubble time (Besla et al 2012; Nichols et al 2011).
This implies substantial ellipticity of the orbit with the Stream
distance over the SGP falling within the range $80-150$ kpc (Model 1;
Besla et al 2012). Given the uncertain mass of the Galactic halo
(Kafle et al 2012), the drag coefficient of the Stream gas, and the
initial orbit parameters of the Magellanic Clouds, the true distance
along the SGP is unlikely to exceed 100 kpc (Jin \& Lynden-Bell 2008).

The Magellanic Stream is made up of a series of dense gas clumps with
column densities that vary over at least a factor of ten (Moore \&
Davis 1994; Putman et al 1998). Even the diffuse gas between the dense
clouds is optically thick to the Lyman continuum ($>1.6\times 10^{17}$
cm$^{-2}$).  In the clouds, a mean column density of $N_c \sim 7\times
10^{19}$ cm$^{-2}$ and a mean cloud size of $d_c \sim 1$ kpc leads to
a total hydrogen density spanning the range $n_H \approx 0.03 - 0.2$
cm$^{-2}$. This leads to a typical spherical cloud mass of roughly
$M_c \sim m_p N_c d_c^2/2 \sim 10^6$ M$_\odot$, for which $m_p$ is the
proton mass. In reality, the gas may have a fractal distribution in
density (Bland-Hawthorn et al 2007; Stanimirovic et al 2008; Nigra et
al 2012).  The mean metallicity of the Magellanic Stream appears to be
everywhere one tenth of the solar value (Fox et al 2013) although
isolated regions close to the Magellanic Clouds are more enriched
(Richter et al 2013).

\begin{figure}[htbp]
   \centering
 \includegraphics[scale=0.45,trim=35mm 10mm -20mm 20mm,
 clip=true]{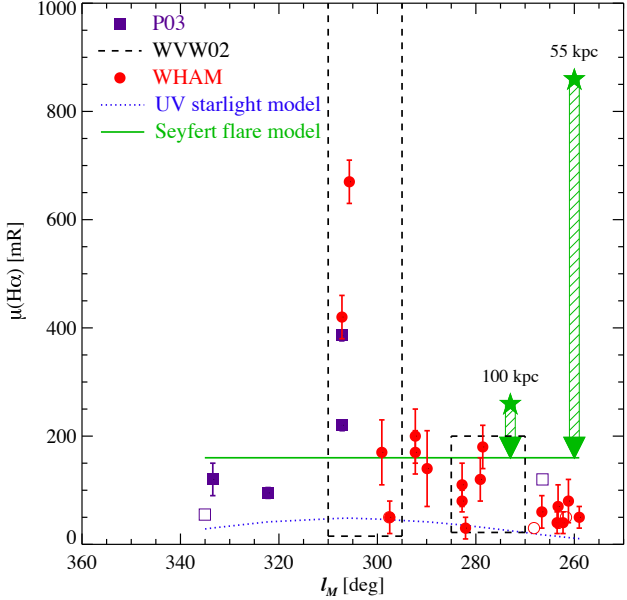} 
   \caption{
Observations and models of H$\alpha$ emission along the Magellanic
Stream.  The dashed boxes indicate the range of detected values from
Weiner et al (2002); the purple points are from Putman et al (2003);
the red points are new observations from WHAM (Madsen 2012). The
extreme H$\alpha$ values occur close to the SGP at $\ell_M \approx
303^\circ$. The dotted blue curve is the upper bound of allowed UV
ionisation from the Galaxy (disk+bulge+hot gas) from the model of
Bland-Hawthorn \& Maloney (1999). The green arrows illustrate the
effect of a fading Seyfert flare ($f_E=0.1$) for Stream distances of 55 kpc (long
arrow) and 100 kpc (short arrow). The horizontal green line indicates
a characteristic H$\alpha$ surface brightness (160 mR). 
     }
   \label{f:em}
   \medskip
\end{figure}

\medskip
\noindent {\sl H$\alpha$ observations.}  
Weiner \& Williams (1996) made the remarkable discovery of relatively
bright H$\alpha$ emission along the Magellanic Stream when compared to
high-velocity clouds (HVC) close to the Galactic plane. These
detections have been confirmed and extended through follow-up
observations (Weiner et al 2002; Putman et al 2003; Madsen 2012) that
are summarised in Fig. 2.  The figure shows the H$\alpha$ surface
brightness observations along the Stream as a function of Magellanic
longitude $\ell_M$, where $\ell_M$ is defined in a plane that lies
close to a great circle passing through the South Galactic Pole
(Nidever 2008). Solid symbols show detections; open symbols show
non-detections. In order to minimize the effects of bright,
time-variable atmospheric emission lines, the data taken with the
Wisconsin H-Alpha Mapper (WHAM) employed an offset sky subtraction
technique (Madsen et al 2001). The high spectral resolution of the
WHAM data enables us to confirm the association of the ionised gas
with the cold Stream gas; the \HI\ and H$\alpha$ velocities are
consistent with each other to within $\approx$ 5 km s$^{-1}$.  The
densities and length scales for the \HI\ clouds derived above are
within range of the expected values to account for the mean H$\alpha$
surface brightness. Furthermore, the beam size for most of the
H$\alpha$ measurements (1$^\circ$ for WHAM) is of order the mean
projected cloud size and thus provides an average estimate for the
cloud. However, the origin of this emission remains highly uncertain.

\medskip
\noindent{\sl Stream ionisation.}
We adopt physically motivated units that relate the ionising photon
flux at a distant cloud to the resultant H$\alpha$ emission.  For
this, we need to relate the plasma column emission rate to a photon
surface brightness. In keeping with astronomical research on diffuse
emission (e.g.  WHAM survey $-$ Reynolds et al 1998), we use the
Rayleigh unit introduced by aeronomers (q.v. Baker \& Romick 1976)
which is a unique measure of {\it photon} intensity; 1 milliRayleigh
(mR) is equivalent to $10^3/4\pi$ photons cm$^{-2}$ s$^{-1}$
sr$^{-1}$.  The emission measure ${\cal E}_m$ for a plasma with
electron density $n_e$ is given by (e.g. Spitzer 1978)
\bee
{\cal E}_m = \int f_i n_e^2\ {\rm d}z \ \ \ \ {\rm cm^{-6} \ pc}
\label{e:em}
\eee
which is an integral of H recombinations along the line of sight $z$
multiplied by a filling factor $f_i$. The suffix $i$ indicates that we
are referring to the volume over which the gas is ionised. For a
plasma at 10$^4$K, ${\cal E}_m(\rm H\alpha)=1$ cm$^{-6}$ pc is
equivalent to an H$\alpha$ surface brightness of 330 milliRayleighs
(mR).  In cgs units, this is equivalent to $1.9\times 10^{-18}$ erg
cm$^{-2}$ s$^{-1}$ arcsec$^{-2}$ which would be a faint spectral
feature in a 1 hr integration using a slit spectrograph on an 8m
telescope. But for the Fabry-Perot `staring' technique employed in
Fig.~\ref{f:em}, this is an easy detection if the diffuse emission
uniformly fills the aperture. We refer to the Stream H$\alpha$
emission as relatively bright because it is much brighter than
expected for an optically thick cloud at a distance of 50 kpc or 
more from the Galactic Centre.

The characteristic H$\alpha$ surface brightness observed along the
Stream of $\mu_{{\rm H}\alpha} \approx 160$ mR (Fig.~\ref{f:em}) can be
used to set a minimum required ionising photon flux and luminosity for a
Galactic Centre flare,
assuming that the Stream emission has not begun to fade and that there
is no absorption or extinction of the ionising photon flux prior to
reaching the Stream. For a slab ionized on one side, this is
\bee
\varphi_{i,{\rm min}} \approx 3.9\times 10^5\;{\rm phot\; cm^{-2}\; s^{-1}}
\eee
The implied ionising photon luminosities are then
\bee
N_{i,{\rm min}} \approx (1.4 - 4.7)\times 10^{53}\;{\rm phot\; s^{-1}}
\eee
for $D= 55 - 100$ kpc. {\it Any model in which the Stream emission is
  produced by a nuclear outburst must explain the magnitude of this
  ionising photon luminosity.}

\medskip
\noindent{\sl Expected emission via the Galactic stellar population.} 
First we consider the expected emission due
to the ionizing radation from the Galactic stellar population and associated sources.
The total flux at a frequency $\nu$ reaching an observer
located at a distance $D$ is obtained from integrating the specific
intensity $I_\nu$ over the surface of a disk, i.e.
\bee
F_\nu = \int_A I_\nu({\bf n})({\bf n}.{\bf N}) {{dA}\over{D^2}}
\label{e:flux}
\eee
where ${\bf n}$ and ${\bf N}$ are the directions of the line of sight
and the outward normal to the surface of the disk, respectively.
At this stage, we consider only an isotropic illumination source
rather than more complex forms of illumination with a strong
polar angle dependence. In this instance,
we use the more familiar scalar form of eq.~(\ref{e:flux}) such that
\bee
\varphi_\star = \int_\nu {{F_\nu}\over{h\nu}} \cos\theta\ d\nu
\label{e:star}
\eee
where $\varphi_\star$ is the photoionising flux from the stellar
population, ${\bf n}.{\bf N} = \cos\theta$ and $h$ is Planck's
constant. This is integrated over frequency from the Lyman limit
($\nu=13.6\; {\rm eV}/h$) to infinity to convert to units of photon
flux (phot cm$^{-2}$ s$^{-1}$). The photon spectrum of the Galaxy is a
complex time-averaged function of energy $N_\star$ (photon rate per
unit energy) such that $4\pi D^2 \varphi_\star = \int_0^{\infty}
N_\star(E)\: dE$.

For a given ionising luminosity, we can determine the expected
H$\alpha$ surface brightness at the distance of the Magellanic Stream.
For an optically thick cloud ionised on one side, we relate the
emission measure to the ionising photon flux using ${\cal E}_m =
1.25\times 10^{-6} \varphi_\star$ cm$^{-6}$ pc (Bland-Hawthorn \&
Maloney 1999).  The total ionising luminosity of the Galaxy is now
well established within a small factor (Bland-Hawthorn \& Maloney
1999; Weiner et al 2002; Putman et al 2003). For a total disk star
formation rate of 1.1$\pm 0.4$ M$_\odot$ yr$^{-1}$ (Robitaille \&
Whitney 2010), the hot young stars produce an integrated photon flux
over the disk of $2.6\times 10^{53}$ phot s$^{-1}$ with
very few photons beyond 50 eV.

The mean vertical opacity of the disk at the Lyman limit is
$\tau_{LL}=2.8\pm 0.4$, equivalent to a vertical escape fraction of
$f_{\star,{\rm esc}} \approx 6$\% perpendicular to the disk (${\bf
n}.{\bf N} = 1$).  The Galactic UV contribution at the distance $D$ of
the Magellanic Stream is given by
\bee
\mu_{\star,{\rm H}\alpha} = 21 \zeta \left({f_{\star,{\rm
      esc}}}\over{0.06} \right) \left({D}\over{{\rm 55\
    kpc}}\right)^{-2} \ \ \ {\rm mR} . 
\label{f:Gal}
\eee
corresponding to $\varphi_i \simeq 5.1\times 10^4$ phot s$^{-1}$.  The
correction factor $\zeta \approx 2$ is included to accommodate weakly
constrained ionising contributions from the old stellar population,
hot gas (disk$+$halo) and the Magellanic Clouds (Slavin et al 2000;
Bland-Hawthorn \& Maloney 2002; Barger et al 2013). We arrive at the
blue curve presented in Fig.~\ref{f:em} which fails to explain the observed
H$\alpha$ emission by at least a factor of 5.

We note that if the Stream is more distant at 100 kpc, preferred by
some recent models (Besla et al 2012), the predicted emission measure
due to the Galaxy approaches the upper limit ($\sim 8$ mR at
2$\sigma$) obtained by Weymann et al (2001). The discrepancy with the
observed H$\alpha$ brightness is now a factor of 20!

\begin{figure}[htbp]
   \centering
   \includegraphics[scale=0.3]{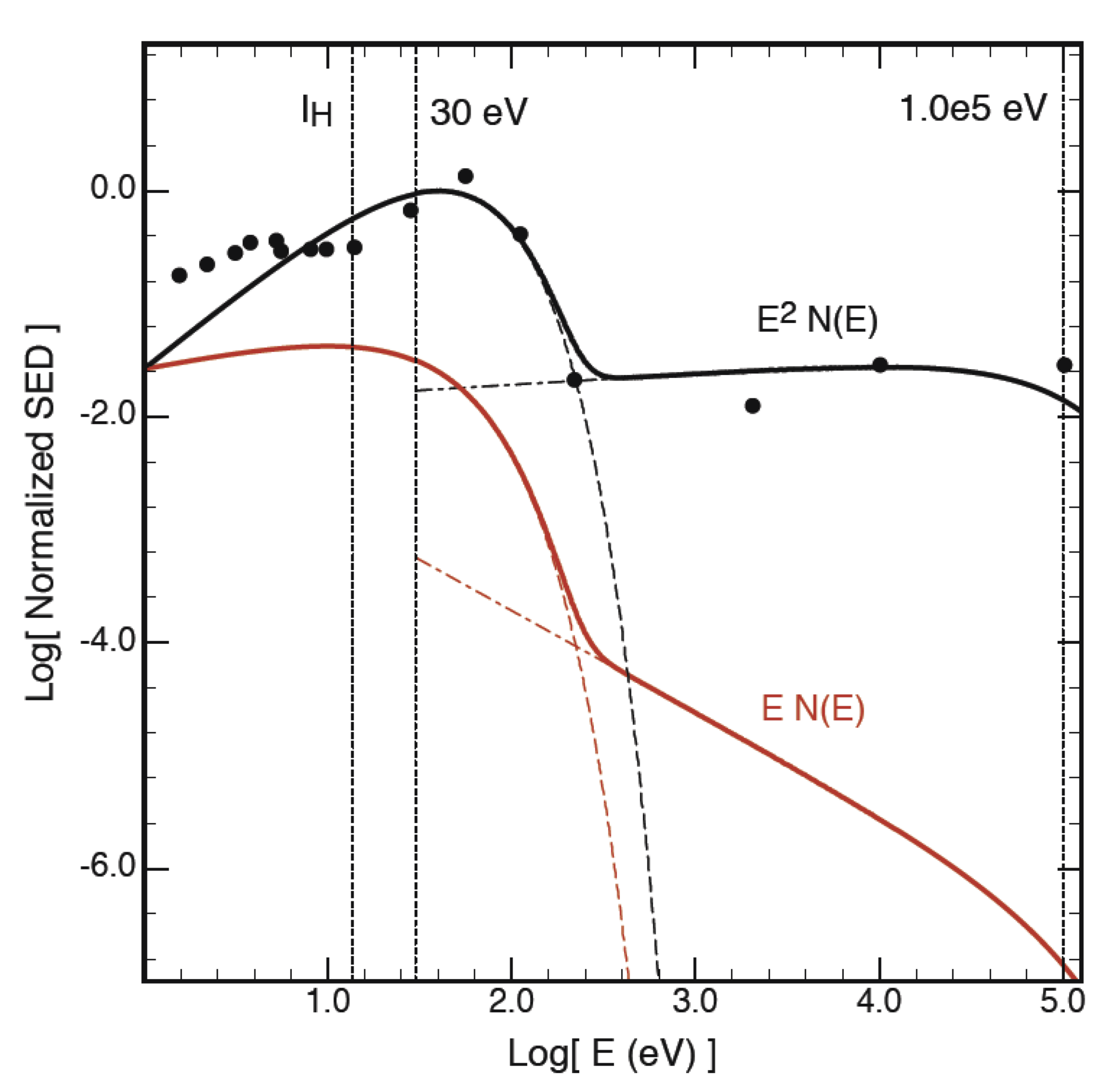}
 \caption{
The energy spectrum of the Galactic Centre accretion disk (see
Appendix C) is assumed to comprise a `big blue bump' and an X-ray $+$ $\gamma$-ray
power law component. The lower curve is the photon spectrum
$E.N_\bullet(E)$. The upper curve is the energy-weighted photon
spectrum $E^2.N_\bullet(E)$ which serves to illustrate that there is
an order of magnitude more energy in the big blue bump ($\eta = 9$)
than the hard energy tail. The data points are taken from the nuclear
ionising spectrum derived from the {\it ISO-SWS} ($2-200\mu $m)
satellite data for NGC 1068.}
   \label{f:agn}
   \medskip
\end{figure}

\medskip
\noindent{\sl Nuclear spectrum.}
We now consider the Stream emission due to ionizing radiation powered by the supermassive
black hole at the Galactic Centre.
Motivated by detailed spectral observations of AGN, we adopt a two-component 
accretion-disk model for the photon spectrum of the central source. We
define the specific photon luminosity for the two-component model by
\bea
N_\bullet &=& k_1 (E/E_1)^{-2/3} e^{-E/E_1} + \cr
& & k_2 (E/E_2)^{-\alpha}
e^{-E/E_2} {\cal H}(E-E_1)
{\quad \rm phot\ s^{-1}\ eV^{-1}}
\label{e:agn}
\eea
where ${\cal H}[E-E_1] = 1$ if $E>E_1$ and ${\cal H}[E-E_1] = 0$
otherwise. 
The first term on the rhs of eq.~(\ref{e:agn}) represents the cool,
optically thin Shakura-Sunyaev spectrum thought to produce the
enhanced UV (``big blue bump'') emission observed in Seyfert galaxies
and quasars (Antonucci 1993). The second term represents the X-ray and
gamma-ray emission observed from the source.  We choose $E_1 = 30$ eV
for the cool outer blackbody spectrum, $\alpha = 1.9$ for the {\it
photon} spectral index of the X-$\gamma$ component (i.e., 0.9 for the
energy spectral index), and adopt $E_2 = 100$ keV (Dermer et al 1997).

By integrating eq.~(\ref{e:agn}) weighted by energy, we derive the
relative normalisation constants $k_1$ and $k_2$ (see Appendix C).  The
ionising spectrum of NGC 1068 is strongly constrained by $2-200\;\mu$m
{\it ISO-SWS} observations (Alexander et al 2000; Lutz et al 2000).
This barred spiral galaxy has the most extensive literature of any Seyfert
and serves as a surrogate for the AGN activity at the Galactic Centre.
We have used the unattenuated spectrum to normalize our composite model.  
We find that $\eta \approx 9$ ($\log k_1/k_2=8.8663$) provides a
reasonable match to NGC~1068. Our assumed ionising spectrum of the
Galaxy's AGN presented in Fig.~\ref{f:agn} has the same functional form.

The ionising flux $\varphi_\bullet$ follows from eq.~(\ref{e:agn}) such
that $4\pi D^2 \varphi_\bullet = \int_{I_H}^\infty N_\bullet dE$ where
$I_H = 13.598$ eV.  The spectrum is dominated by the soft component
such that estimates of the AGN-induced H$\alpha$ emission (below) are
not obfuscated by the longer mean free paths of the harder
photons. Thus properties derived from the stellar ($\varphi_\star$)
and AGN ($\varphi_\bullet$) photon fluxes can be compared directly;
the photons propagate roughly the same depth into an \HI\ cloud.

\medskip
\noindent{\sl Expected emission from an active nucleus.}
We now relate the accretion disk luminosity $L_\bullet$ to the
properties of the supermassive black hole. An accreting black hole
converts rest-mass energy with a conservative efficiency
$\epsilon=5\%$ into radiation with a luminosity ($=\epsilon \dot m
c^2$)
\bee
L_\bullet \approx 7.3\times 10^{11}\left({\epsilon\over
  0.05}\right)\left({\dot m}\over {M_\odot\; {\rm yr}^{-1}}\right)\;L_\odot
\label{e:acclum}
\eee
for which $\dot m$ is the mass accretion rate.  The accretion disk
luminosity can limit the accretion rate through radiation
pressure. The so-called Eddington limit is given by

\bea
L_E &=& {{4\pi G M_\bullet m_p c}\over{\sigma_T}} \\
                    &=& 1.4\times 10^{11}\left({M_\bullet\over 4\times
                    10^6\;M_\odot}\right)\;L_\odot 
 \label{e:eddlum}
\eea
where $M_\bullet$ is the black-hole mass and $\sigma_T$ is the Thomson
cross-section for electron scattering.

Radiation pressure from the accretion disk at the Galactic Centre
limits the maximum accretion rate to $\dot m \sim 0.2$ M$_\odot$
yr$^{-1}$.  Some Seyferts, including NGC 1068 (Begelman \&
Bland-Hawthorn 1997), radiate at close to the limit. But active
galactic nuclei appear to spend most of their lives operating at a
fraction $f_E$ of the Eddington limit with rare bursts arising from
accretion events (Hopkins \& Hernquist 2006).  For clouds at a
distance of 55$-$100 kpc along the SGP, we now show that only a fraction of
the maximum accretion rate is needed to account for the H$\alpha$
emission along the Magellanic Stream.  The orbital period of the Stream
is of order a Hubble time so we can consider the Stream to be a
stationary target relative to ionisation timescales.

The dust levels are very low in the Stream (Fox et al 2013); internal
and line-of-sight extinctions are negligible. It follows from
eq.~(\ref{e:eddlum}) that the ionising photon luminosity of the Seyfert
nucleus is given by
\bee
N_{\bullet,i} =  7.2\times
10^{53}\left({{f_E}\over{0.1}}\right)\left({M_\bullet\over 4\times
  10^6\;M_\odot}\right)\;{\rm phot\; s^{-1} } . 
\label{e:ion}
\eee
For the photon spectrum in eq.~(\ref{e:agn}), we find that 20\% of the energy
falls below the Lyman limit and therefore does not photoionise hydrogen.
If the absorbing cloud is optically thick, the
ionising flux can be related directly to an H$\alpha$ surface
brightness. The ionising flux is given by
\bee
\varphi_\bullet = 2.0\times 10^6 \left({{f_E}\over{0.1}}\right)
\left({f_{\bullet,{\rm esc}}}\over{1.0} \right)\left(D\over{\rm 55\
  kpc}\right)^{-2} \ \ \ {\rm phot\ cm}^{-2}\ {\rm s}^{-1} . 
\label{e:phi_agn}
\eee
We have included a term for the UV escape fraction from the AGN
accretion disk $f_{\bullet,{\rm esc}}$ (${\bf n}.{\bf N}=1$). This is
likely to be of order unity to explain the integrated energy in
observed ionisation cones (Sokolowski et al 1991; Mulchaey et al
1996). Some energy is lost due to Thomson scattering but this is known
to be only a few percent in the best constrained sources (e.g. NGC
1068; Krolik \& Begelman 1986).  In principle, the high value of
$f_{\bullet,{\rm esc}}$ can increase $f_{\star,{\rm esc}}$ but the
stellar bulge is not expected to make more than a 10-20\% contribution
to the total stellar budget (Bland-Hawthorn \& Maloney 2002); a
possible contribution is accommodated by the factor $\zeta$
(eq.~[\ref{f:Gal}]).

The expected surface brightness for clouds that lie within a putative
`ionisation cone' from the Galactic Centre is given by
\bee
\mu_{\bullet,{\rm H}\alpha} = 825 \; b\; \left({{f_E}\over{0.1}}\right)
\left({f_{\bullet,{\rm esc}}}\over{1.0} \right)\left(D\over{\rm 55\
  kpc}\right)^{-2} \ \ \ {\rm mR} . 
\label{e:mu_agn}
\eee
Strictly speaking, this provides us with an upper limit or `peak
brightness.' In \S 3, we show that proper consideration must be given
to the physical state of the gas and the time since the event
occurred. The recombination emission will fade once the burst duration
has passed and the gas begins to recombine and cool.
For completeness, we have included a beaming factor $b$ to accommodate
more exotic models that allow for mild beaming of the UV radiation
(e.g. Acosta-Pulido et al 1990).  The solid angle subtended by a
half-opening angle of $\theta_{1/2}$ is $\Delta\Omega =
2\pi(1-\cos\theta_{1/2})$. So the beam factor $b =
(1-\cos\theta_{1/2})^{-1}$ expresses how much of the isotropic
radiation is channelled into a cone rather than $2\pi$ sr. For
example, $\theta_{1/2} = 22.5^\circ$ is a beam factor $b = 13$;
$\theta_{1/2} = 30^\circ$ is a beam factor $b = 7.5$; $\theta_{1/2} =
90^\circ$ is a beam factor $b = 1$ (isotropic emission) adopted for
the remainder of the paper. The emission within an ionisation cone can
be isotropic if the restriction is caused by an external screen,
e.g. a dusty torus on scales much larger than the accretion disk.

\begin{figure*}[htbp]
   \centering \includegraphics[scale=0.5,trim=0mm 0mm 0mm
  0mm,clip=true]{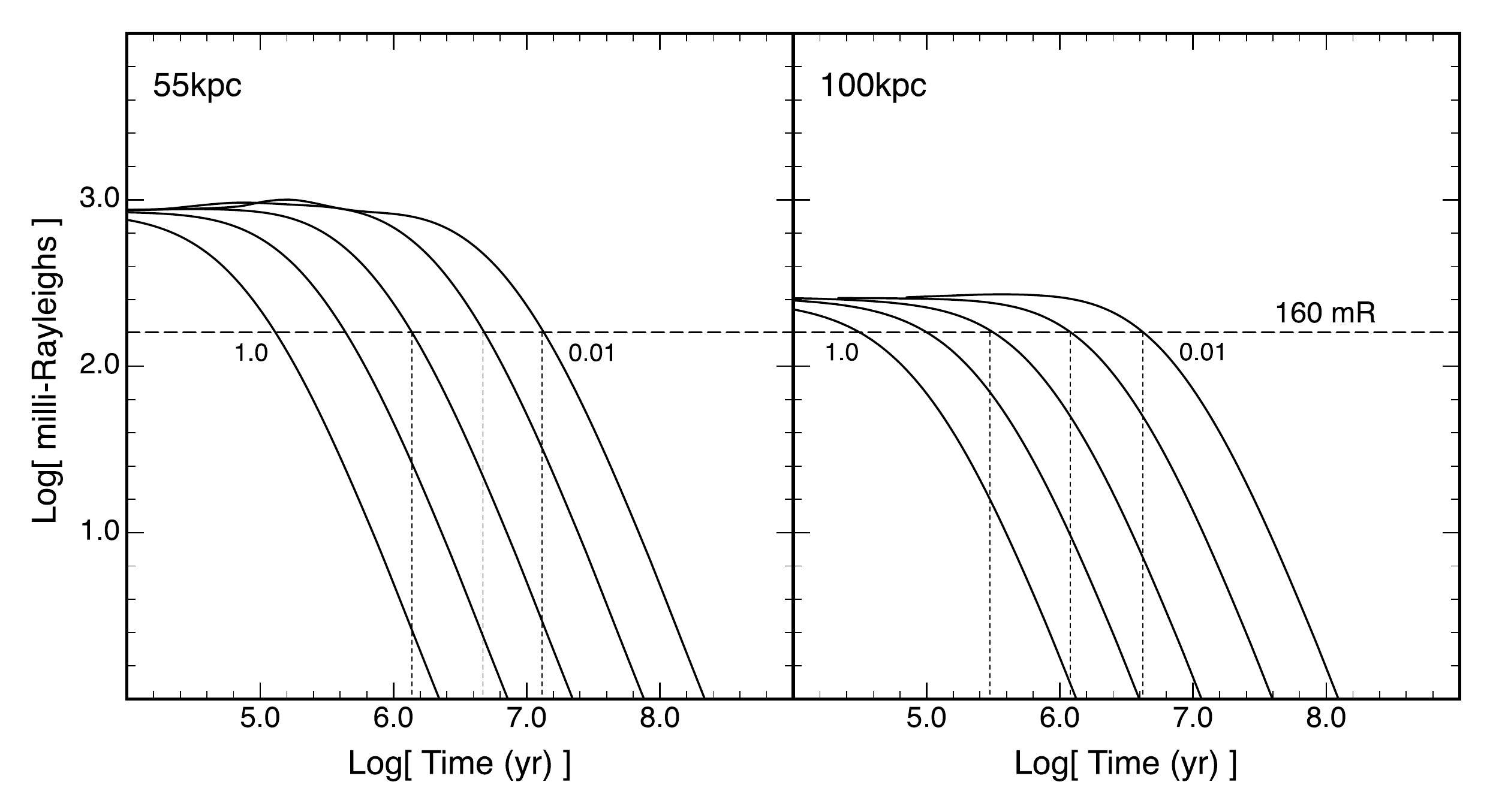}
  \caption{{\it MAPPINGS IV} time-dependent isochoric calculations of
  the change in H$\alpha$ surface brightness after a Seyfert flare has
  occurred at the Galactic Centre.  {\it (Left)} $D=55$ kpc,
  $f_E=0.1$; {\it (Right)} $D=100$ kpc, $f_E=0.1$.  From left to
  right, the pre-ionised gas densities are 1.0, 0.3, 0.1, 0.03, 0.01
  cm$^{-3}$. The clouds have 0.1 Z$_\odot$ metallicity and are
  irradiated by an AGN accretion disk (see text). The mean surface
  brightness of the Stream near the SGP (160 mR) is shown as a
  horizontal dashed line. Physical properties of the models are to be
  found in Table 1.}
   \label{f:map_Em}
   \medskip
\end{figure*}

\section{Past AGN activity}

\subsection{Timescales}
Consider the situation in which we observe the ionisation of the
Stream due to a nuclear source. A burst of intense UV at a time $T_o$
in the past, lasting for a period $T_B$, must propagate for a time $T_C
\sim 0.18\ (D/55\ {\rm kpc})$ Myr to reach the Stream. For example,
in the Seyfert jet model of Guo \& Mathews (2012), $T_o \approx 1-3$
Myr and $T_B \approx 0.1-0.5$ Myr. The ionisation front then moves
through the cool gas until the UV is used up. This occurs in a time
$T_I \sim 1/(\sigma_H \varphi_\bullet)$ where $\sigma_H$ is the H
ionisation cross section. This follows from the fact that the speed of
the ionisation front into the neutral gas is $q =\varphi_\bullet /
n_H$ where $q$ is the ionisation parameter\footnote{This is often
defined as the dimensionless ionisation parameter $u=q/c$.} for a
neutral H density $n_H$. For the H$\alpha$ emission levels associated
with the Magellanic Stream, $T_I \sim 4\times
10^3/(\varphi_\bullet/10^6)$ yr; for simplicity, the extra factors
from eq.~(\ref{e:phi_agn}) are not carried over. The time for the gas to
reach ionisation equilibrium will be of order $T_I$, and is very short
compared to both $T_C$ and the likely values of $T_o$. The H$\alpha$
emission is then proportional to $\alpha_B n_e N_e$, where $n_e$ is
the electron density, $N_e$ is the column density of ionised gas, and
$\alpha_B$ is the Case B recombination coefficient.

Since the level of activity in the Galactic Center today is far too
low to produce the observed H$\alpha$ emission, the Stream emission in
this picture is almost certainly fading from an earlier peak. The time
$T_d$ required for the emission to decrease from its peak value to the
observed brightness depends on both the time evolution of the burst
luminosity and the gas density in the Stream\footnote{This is
discussed in detail in \S 5.1 and Appendix A.}. Since we observe the
Stream as it was $T_C$ years ago, the look-back time to the initial
event is $T_o = T_d + 2T_C$ (assuming the burst time $T_B$ is much
less than $T_o$). We now revisit these approximations with detailed
calculations of the ionisation state of the gas.

\begin{figure*}[htbp]
   \centering
  \includegraphics[scale=0.5,trim=0mm 0mm 0mm 0mm,clip=true]{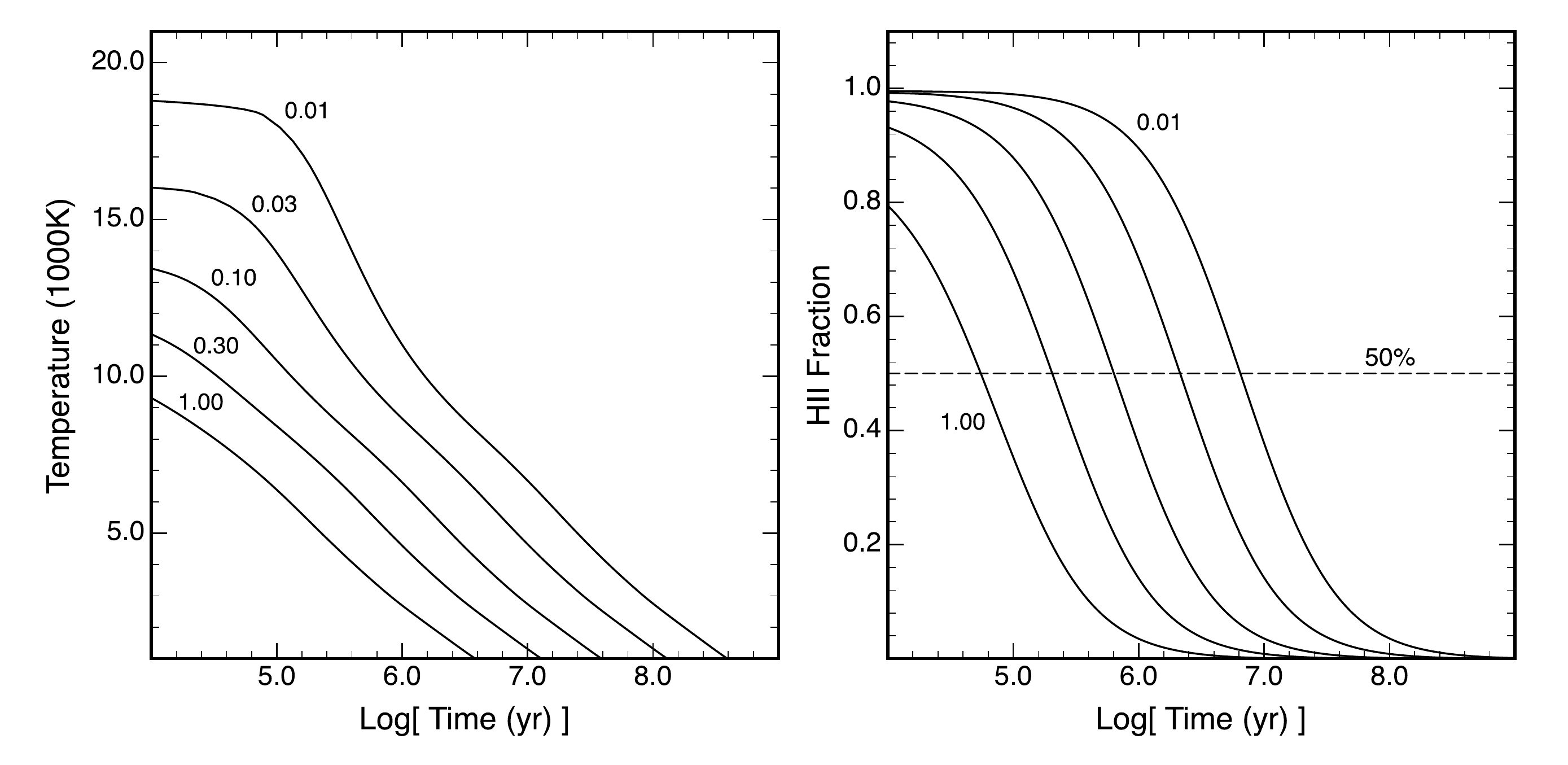}
\caption{{\it MAPPINGS IV} time-dependent calculations of how the
   ionised surface of a dense gas clouds cools with time: {\it (Left)}
   electron temperature $T_e$; {\it (Right)} ionisation fraction
   $\chi$. We show the isochoric (constant density) models where the
   densities in cm$^{-3}$ are indicated; isobaric models cool twice as
   fast at late times. The \HI\ clouds have 0.1 Z$_\odot$ metallicity
   and are irradiated at $D=55$ kpc by an AGN accretion disk ($f_E =
   0.1$) at the Galactic Centre (Fig. 3).}
   \label{f:map_T}
   \medskip
\end{figure*}

\subsection{UV photoionisation}

We use the recently completed {\it MAPPINGS IV} code (Dopita et al
2013) to study both isochoric and isobaric cooling at the surface of
the Magellanic Stream. The source of the impinging radiation field is
the accretion disk model presented in Fig.~\ref{f:agn}. The
models were run by turning on the source of
ionization, waiting for the gas to reach ionization/thermal equilibrium,
and then by turning off the ionizing photon flux.  In
Figs.~\ref{f:map_Em} and \ref{f:map_T}, we present our modelled trends
in gas temperature ($T_e$), ionisation fraction ($\chi$) and emission
measure (${\cal E}_m$) for time-dependent ionisation of the Magellanic
stream. Important properties of the medium $-$ ionised column depths,
emission measures, cooling times, etc. $-$ are included in
Table~\ref{t:map}. The sound crossing times of the warm ionised layers
are too long ($\gta$10 Myr) in the low density regime relevant to our
study for isobaric conditions to prevail. Both a near
($D=55$ kpc) and a far ($D=100$ kpc) distance is considered.

The gas phase abundances are the solar values scaled to [Fe/H] $=$ -1
now well established from {\it HST COS} measurements (Fox et al 2013;
Richter et al 2013).  The upper limits for [OI]630 nm from the WHAM
survey indicate that the ionisation fraction in the brightest
H$\alpha$-emitting clouds exceeds $\chi = 50$\% (Madsen et al 2013).
For a spherical cloud, its mass is approximately $M_c \sim f_n\rho_c
d_c^3/2$ where the subscript $n$ denotes that the filling factor
refers to the neutral cloud prior to external ionisation.  For a fixed
cloud mass (or equivalently, cloud column density $N_c$), the filling
factor is inversely related to the \HI\ gas density. Any value other
than unity leads to higher gas densities, and shorter recombination
times.  While the cloud geometry and the volume filling factor are
uncertain (Fox et al 2010), self-consistent ionisation parameters $q$
are obtained at all times and lie within the range $\log q = 5.6-7.6$
($\log u \approx$ -5 to -3).

In Fig.~\ref{f:map_T}, the initial photoionising flash rapidly heats
the gas to a peak temperature before the gas begins recombining at a rate that is
inversely proportional to $n_H$ as expected (\S 2). We observe that
the recombination rate is faster than the cooling rate during this
period. The initial flash produces high ionisation states
(e.g. \HeII, \OIII\ emission lines) but these fade rapidly. In
Fig.~\ref{f:map_Em}, the increase in electron density leads to a
maximum H$\alpha$ brightness which then also fades. The Balmer
decrement H$\alpha$/H$\beta$ is everywhere in the range $3.0-3.1$
until very late times when it begins to climb, except now the flare
ionisation signal has almost faded from view. While the decrement
is sensitive to dust extinction, the low metallicity of the Stream has
negligible impact on this diagnostic (see \S 5). At late times, the
H$\alpha$ surface brightness in all cases scales as $t^{-2}$; see
Appendix A.

\begin{table}[htdp]
\caption{MAPPINGS IV time-dependent ionisation
  calculations\footnote{Seyfert flare model ($f_E=0.1$) using two
  distances ($D=$ 55, 100 kpc) for the Magellanic Stream. The columns
  are: (1) hydrogen gas density; (2) depth of ionised layer integrated
  to 90\% neutral; (3) H$\alpha$ surface brightness in erg cm$^{-2}$
  s$^{-1}$ arcsec$^{-2}$; (4) H$\alpha$ surface brightness in mR; (5)
  time for the ionised gas to cool down to ${\cal E}_m = 160$ mR; (6)
  look-back time $T_o$ $=$ $T_R$ $+$ twice the light propagation time
  ($T_C$).  Emboldened rows are consistent with the known gas
  properties of the Stream.}}
\begin{center}
\begin{tabular}{lllcll}
& & (a) 55 kpc & & \\
\\
$n_H$(cm$^{-3}$)  & $d_m$(pc) & $\mu_{{\rm H}\alpha}$(cgs) &
$\mu_{{\rm H}\alpha}$(mR) & $T_R$(yr) & $T_o$(yr)\\ 
\\
1.0    & 9          & 4.8e-18 & 844 & 1.3e5 & 4.9e5 \\ 
{\bf 0.3}  & {\bf 63}        & {\bf 4.8e-18} & {\bf 848} & {\bf 4.3e5}
& {\bf 7.9e5} \\ 
{\bf 0.1}  & {\bf 404}      & {\bf 4.8e-18} & {\bf 849} & {\bf 1.4e6}
& {\bf 1.8e6} \\ 
{\bf 0.08} & {\bf 1423}   & {\bf 4.8e-18} & {\bf 852} & {\bf 1.8e6} &
{\bf 2.9e6} \\ 
0.03  & 3461    & 4.9e-18 & 858 & 4.7e6 & 5.1e6 \\ 
\\
& & (b) 100 kpc & & \\
\\
$n_H$(cm$^{-3}$)  & $d_m$(pc) & $\mu_{{\rm H}\alpha}$(cgs) &
$\mu_{{\rm H}\alpha}$(mR) & $T_R$(yr) & $T_o$(yr)\\  
\\
1.0    & 4          & 1.4e-18 & 251 & 3.0e4 & 7.5e5 \\
{\bf 0.3}    & {\bf 31}        & {\bf 1.5e-18} & {\bf 258} & {\bf
  1.0e5} & {\bf 8.2e5} \\ 
{\bf 0.1}    & {\bf 178}      & {\bf 1.5e-18} & {\bf 258} & {\bf
  3.2e5} & {\bf 1.0e6} \\ 
{\bf 0.03}  & {\bf 1345}    & {\bf 1.5e-18} & {\bf 257} & {\bf 1.2e6}
& {\bf 1.9e6} \\  
0.01  & 9230    & 1.5e-18 & 259 & 4.2e6 & 4.9e6 \\ 
\end{tabular}
\end{center}
\label{t:map}
\end{table}

In Table~\ref{t:map}, we show key properties of the models as a
function of the pre-ionisation gas densities. The realistic cases are
shown as emboldened values where we have ignored small factor
uncertainties; the remaining models exceed the properties (either
local or column density) of the Stream clouds. Column 5 shows the
times $T_d$ for the emission to drop to the canonical surface
brightness of 160 mR (see Fig.~\ref{f:em}). The look-back times are
shown in Column 6; note that the light propagation time ($2T_C$)
dominates the high density extremes. These timescales are in line with
published models of the Fermi bubbles (\S 1). The lower ionising flux
at the 100 kpc distance (since the peak luminosity in the models is
fixed) leads to shorter look-back times because the
gas requires less time to reach 160 mR.

\smallskip
\noindent{\sl Distance vs. Flare Luminosity.} It is clear from
Fig.~\ref{f:map_Em} that lower ionising fluxes or larger Stream
distances lead to shorter inferred timescales for the Seyfert flare
event. To within a small factor, the ionisation model (D,$f_E$) = (55
kpc, 0.1) is equivalent to the (100 kpc, 0.3) model once all
timescales are considered. Conversely, the ionisation model (D,$f_E$)
= (100 kpc, 0.1) is equivalent to the (55 kpc, 0.03) model. In
Appendix A, we present a simplified model for the evolution of the
ionisation fraction and H$\alpha$ surface brightness from the Stream
which is in good agreement with the MAPPINGS models shown in
Fig.~\ref{f:map_Em}, which we use to discuss in more detail the
constraints that can be placed on the flare energetics and evolution
in \S 5.1.

\begin{figure}[htbp]
   \centering
  \includegraphics[scale=0.3]{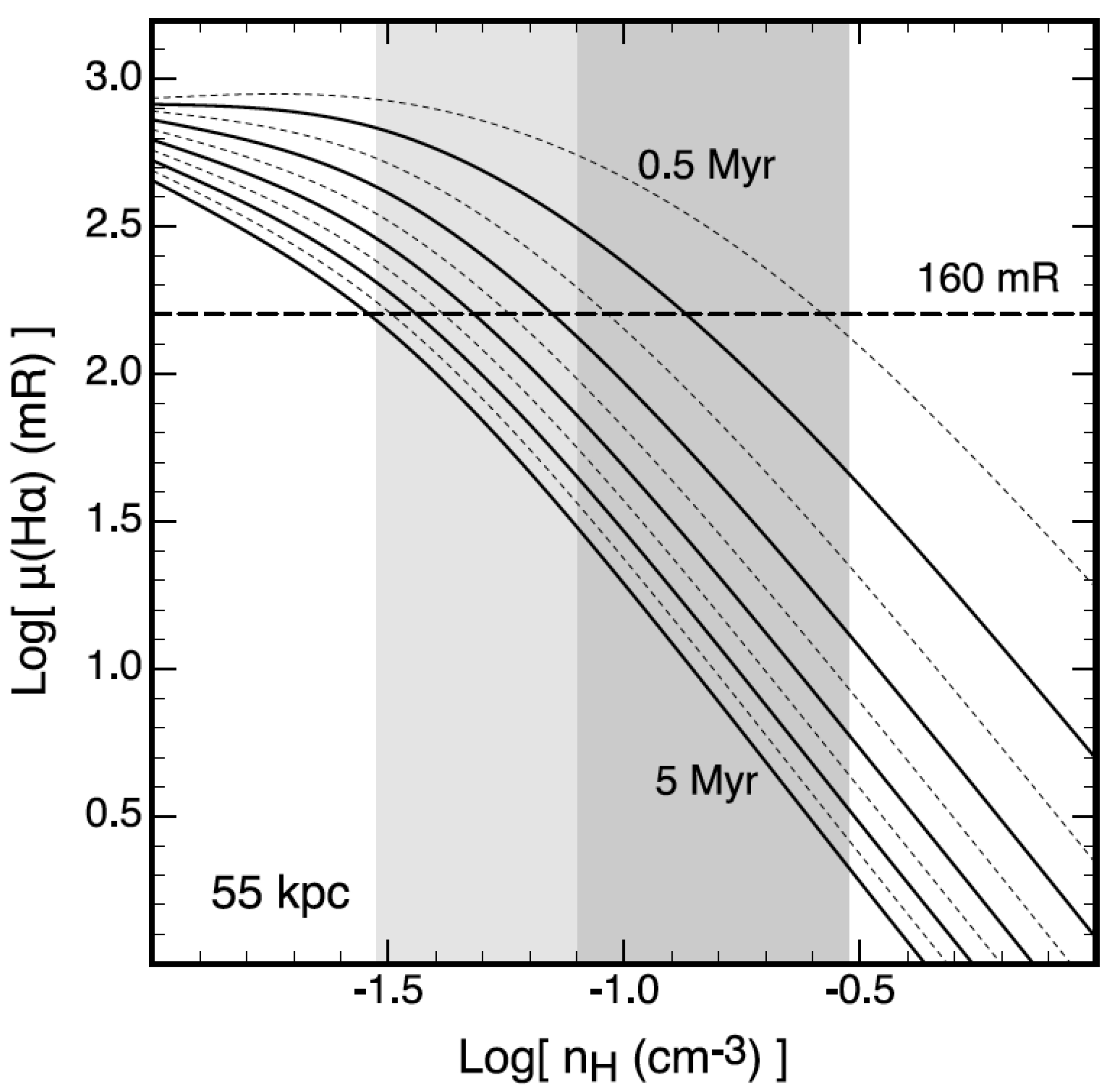}
  \caption{{\it MAPPINGS IV} time-dependent isochoric calculations of
  the change in H$\alpha$ surface brightness after a Seyfert flare has
  occurred at the Galactic Centre ($D=55$ kpc, $f_E=0.1$; Z $=$ 0.1
  Z$_\odot$). The canonical Stream brightness (160 mR) is shown as a
  horizontal dashed line. The tracks plotted every 0.5 Myr show the
  relation between H$\alpha$ and gas density at a fixed recombination
  time $T_R$. The associated look-back times (offset by the light
  crossing time) are shown in Table 1. The dark vertical band shows
  the range in $n_H$ consistent with the known cloud properties; the
  lighter band is marginally consistent. }
   \label{f:fading}
   \medskip
\end{figure}

\smallskip
\noindent{\sl Variations in H$\alpha$ brightness.} An attractive aspect of 
the Seyfert flare model is the ability to accommodate the brightest H$\alpha$
measurements and the scatter about the elevated mean surface
brightness (160 mR) compared to the expective Galactic ionisation
level. An interesting question is whether the scatter reflects
variations in gas density (geometry) or photon arrival time (finite T$_B$). In
Fig.~\ref{f:fading}, we show the relation between $\mu_{{\rm
H}\alpha}$ and $n_H$ at a fixed time over a range of times from 0.5
Myr to 5 Myr. For an impulsive burst, it is possible to accommodate
all of the detections at a fixed time, certainly within the first few
Myr of the Seyfert event. But the very short ionisation timescale
$T_I$ (\S3.1) means that we are unlikely to see temporal variations of
the source: the Stream emission is unaffected by any variations in
incident ionising flux that occurred longer than $\sim T_I$
ago. Source luminosity variations on longer timescales will be
modulated by a transfer function that depends on the gas density (see
Appendix A); any observable fluctuations in H$\alpha$ are likely the
result of variations in density and hence recombination timescale. The
most relevant epoch in interpreting the Stream emission is the end of
the flare.

From the last column in Table 1, {\it we conclude that 
a powerful flare event occurred 1$-$3 Myr ago.}
The inferred magnitude of the event depends on a number of
factors, in particular, the poorly constrained distance to the Magellanic
Stream (see \S 5 and Appendix A), but it must have been at levels
associated with the most energetic Seyfert activity ($f_E > 0.01$).

\section{Future tests of the model}

In Seyferts with moderately low mass black holes, the jet/wind/cone
axis can be strongly misaligned with the spin axis of the galaxy
(Cecil 1988; Mulchaey et al 1996) but there are many counter examples
(Duric et al 1983; Wehrle \& Morris 1987, 1988; Keel et al 2006).  The
Fermi bubbles and the X-ray bipolar structure are roughly aligned with
the SGP.  These features fill most of the conic volume in our model
within 10 kpc, and presumably the outflow has swept any halo
gas aside. We assume that the ionisation cone in the Seyfert
flare model is also aligned with the SGP ($\ell_M \approx 303^\circ$).
Thus to account for the brightest clouds in the Stream
(Fig.~\ref{f:em}), the half-opening angle $\theta_{1/2}$ of the cone
is at least 25$^\circ$ to accommodate enhanced emission at the same
angle from the SGP ($\ell_M \approx 278^\circ$).

Any gas clouds caught within the cones at smaller distances will be
roasted by the Seyfert nucleus. However, almost all the known high
velocity clouds reside close to the Galactic Plane ($b<30^\circ$;
Putman et al 2012). There are few known HVCs close to the SGP although
evidence for ionised HVCs has been presented (Lehner \& Howk 2010).
In our model, most of the HVCs will be fully ionised within the
ionisation cone.  Interestingly, there is one sight line close to
$\ell_M = 308^\circ$ where the H$\alpha$ surface brightness is up to 4
times higher than our benchmark flare value of 160 mR
(Fig.~\ref{f:em}). A possible explanation is that some of the Stream
clouds are somewhat closer than the canonical distance of 55 kpc.

A competing model for the H$\alpha$ emission uses a radiative
hydrodynamic simulation to demonstrate the possibility of a slow shock
cascade acting along the Stream (Bland-Hawthorn et al 2007). Arguably,
this is the {\it only} serious attempt to date to explain the Stream
H$\alpha$ emission. But this model {\it does not} work well if the
Stream at the SGP is at the larger distance of $D\approx 100$
kpc. Given that one end (front) of the Stream is tied to the LMC-SMC
system, this would require the far trailing end (back) of the Stream
to subtend a large angle to the halo, being more radial than
tangential to the halo. This has two problems: (i) the H$\alpha$
emission would be almost entirely confined to the front of the Stream;
(ii) the back of the Stream would be undetectable.

For the near distance of $D=55$ kpc, for most optical diagnostics, the
slow shock cascade may be difficult to disentangle or distinguish from
our model of AGN photoionisation. The diluteness of the predicted AGN
field, with ionisation parameters in the range $\log q = 5.6-7.6$
($\log u \approx$ -5 to -3), tends to produce shock-like emission line
diagnostics.  The high energy part of the big blue bump ($50-100$ eV)
can excite \HeII\ and \OIII, with enhanced ratios to H$\beta$ of about
0.3 and 1 respectively.  But these occur at the peak of the flash and
fade rapidly, and only for the near-field Stream
(Fig.~\ref{f:map_Em}$a$).  The high energy tail in eq.~(\ref{e:agn}) can
excite a few atoms with high ionisation cross sections but the
radiation field in X-rays is very dilute, and the metal fraction is
low ([Fe/H]$\approx$-1; Fox et al 2013).

We are presently re-running the shock cascade models at higher
resolution and with the updated ionisation diagnostics in {\it
MAPPINGS IV}. This will be the focus of a later paper. The shock
cascade has a slightly elevated density-weighted temperature ($T_e
\gta 12,000$ K) compared to the time-averaged Seyfert flare model
($T_e \approx 10,000$ K).  But both models produce comparable emission
in the optical diagnostic \SII, \NII\ and \OI\ emission lines.

A promising diagnostic is the Balmer decrement H$\alpha$/H$\beta$
which is typically enhanced in slow shock models (Bland-Hawthorn et al
2007). The dust content in the Magellanic Stream has negligible impact
on this line ratio. Diffuse optical detection surveys to date have
largely focussed on the 500$-$700nm window in part because H$\beta$ is
harder to detect along most of the Stream (Reynolds et al 1998).  In
our Seyfert flare models, the Balmer decrement at the distance of the
Stream is in the range 3.0$-$3.1 for detectable emission, rising
slowly at late times when the recombination emission has largely
faded. In the shock cascade model, the Balmer decrement exceeds 3.1
and can reach values that are 50\% higher.

\medskip

\section{New insights on AGN activity}
\subsection{Accretion disk}
Sgr A$^\star$ provides us with a front-row seat on the daily life of a
supermassive black hole\footnote{http://swift-sgra.com provides
regular updates on energetic episodes at the Galactic Centre. At the
time of writing, much interest has been sparked by the anticipated
``G2 cloud'' collision -- a warm cloud of several Earth masses --
expected to occur in 2014 (Gillessen et al 2013).}. Nuclear activity
today at the Galactic Centre is remarkably quiescent given the rich
supply of unstable gas within the circumnuclear disk (Requena-Torres
et al 2012). This observation has driven the rapid development of
accretion disk models over the past twenty years: a comprehensive
review is given by Genzel et al (2010). It is now believed that Sgr
A$^\star$ is a radiatively inefficient accretion flow (RIAF) fuelled
by poor angular momentum transport at all radii in part due to strong
outflows, magnetic fields and convection in the innermost accretion zone 
(Blandford \& Begelman 1999; Hawley \& Balbus 2002; cf. Jolley \& Kuncic 2008).  

The observed material within a few parsecs of Sgr A$^\star$
can readily account for the $0.02-0.2$ M$_\odot$ yr$^{-1}$
accretion rate required in our model. Stellar accretion
events are expected once every 40,000 years on average (Freitag et al
2006). There are indications of infalling gas clouds over the past
10 Myr.  One such cloud impact possibly triggered the formation of a kinematically distinct $\sim 10^4$ M$_\odot$ cluster within $\sim 0.1$ pc of Sgr A$^*$, traced by 
$\sim 80$ massive young stars with ages in the range 2.5 -- 8 Myr (Paumard et al 
2006; Lu et al 2013). Wardle \& Yusef-Zadeh (2008) draw
attention to the ``$+$50 km s$^{-1}$ cloud'' known to have 
passed through the Galactic Center within the last 1 Myr. These 
events bracket our inferred epoch for the Seyfert flare which may have
been causally linked to one of these or a related event.

We do not know what the peak luminosity of the AGN burst was, or the
timescale on which it decayed. Such information would shed light on the
nature of the accretion event, whether an individual star ($T_B \sim 10^{2-3}$ 
yr) or an infalling cloud on much longer timescales.
However, we can use the simplified
model for the Stream emission developed in Appendix A, which is in
good agreement with the detailed MAPPINGS IV results for the
time-dependent H$\alpha$ surface brightness, to place constraints on
these quantities.

As in Appendix A, define $\rho$ to be the ratio of the observed
$\mu_{{\rm H}\alpha}$ to its peak value. This is also equal to the
ratio of the minimum required ionizing flux (eq.~[2]) or ionizing
luminosity (eq.~[3]), to their peak values, as well as the
minimum required value of the Eddington fraction $f_{E,{\rm min}}$ to its
peak during the burst. Using eq.~(2) and eq.~(\ref{e:phi_agn}) for the
ionizing flux from the AGN (with $f_{\bullet,{\rm esc}}$ set to 1), we
can write this minimum Eddington fraction as
\bee
f_{E,{\rm min}} = 0.02 \left({D\over 55\;{\rm kpc}}\right)^2
\eee
In the limit where the $e$-folding time for decay of the burst
$\tau_s$ is much shorter than the recombination timescale $\tau_{\rm
  rec}$, we have the analytic result
\bee
f_{E,{\rm peak}} = f_{E,{\rm min}}\left(1+\tau_o\right)^2
\eee
where $\tau_o$ is the dimensionless age of the burst as measured in
recombination times. (This is simply another form of eq.~
[A23]). Using eq.~(A9) for $\tau_{\rm rec}$ and eq.~(14) for $f_{E,{\rm
min}}$, we have calculated the required peak value of $f_E$ as a
function of gas density $n_H$ and burst age $T_o$, for both $D=55$ kpc
and 100 kpc. Assuming that the present-day Eddington fraction of Sgr
A$^\star$ is $\sim 10^{-8}$ (Genzel et al 2010), we can also calculate
the required value of the $e$-folding time $\tau_s$: the inferred
value of peak $f_E$ at a given $T_o$ determines the number of
$e$-folding times that have passed in the age of the burst.

\begin{figure*}[htbp]
  \includegraphics[scale=0.55,clip=true,trim=15mm 0mm 0mm 0mm]{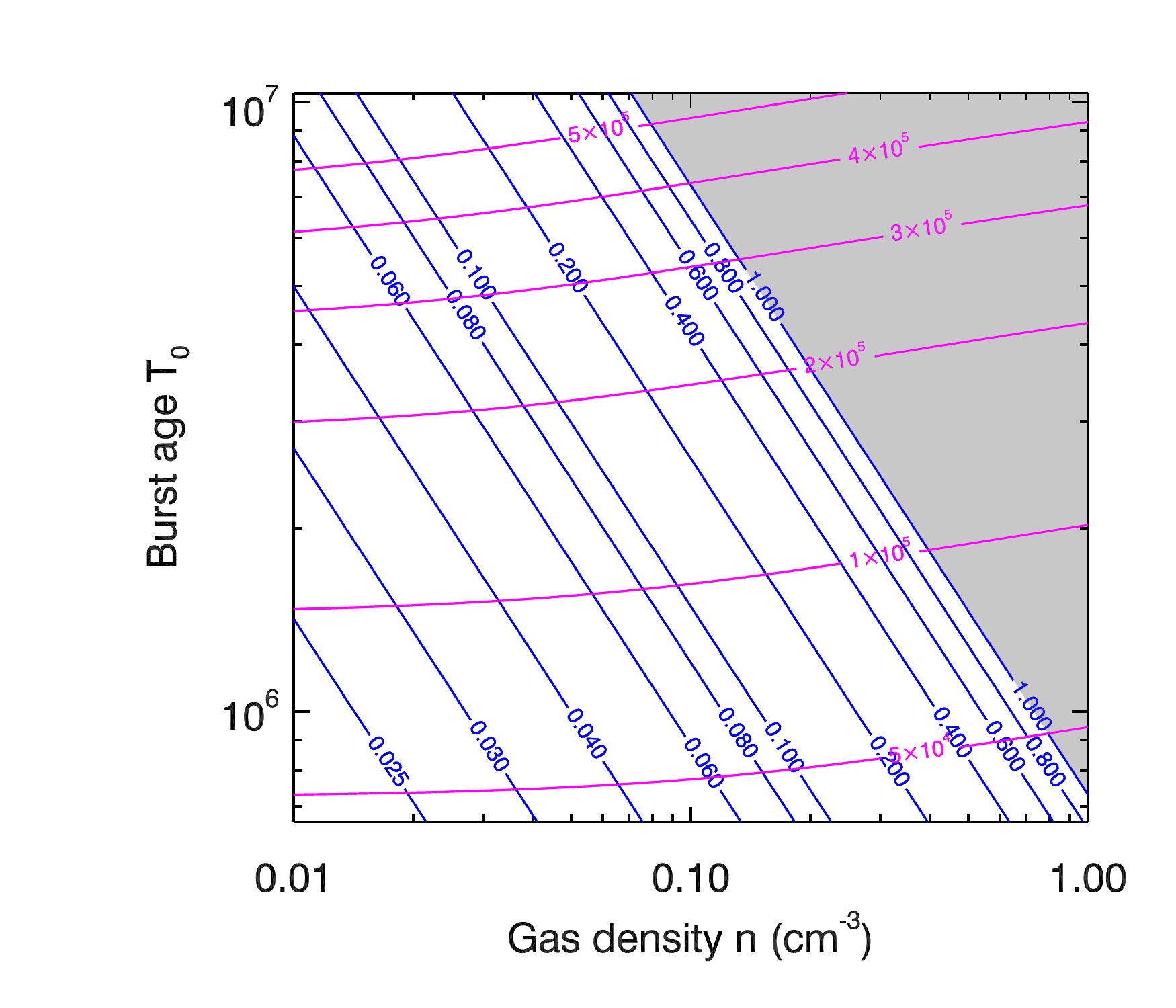}
  \includegraphics[scale=0.55,clip=true,trim=15mm 0mm 0mm 0mm]{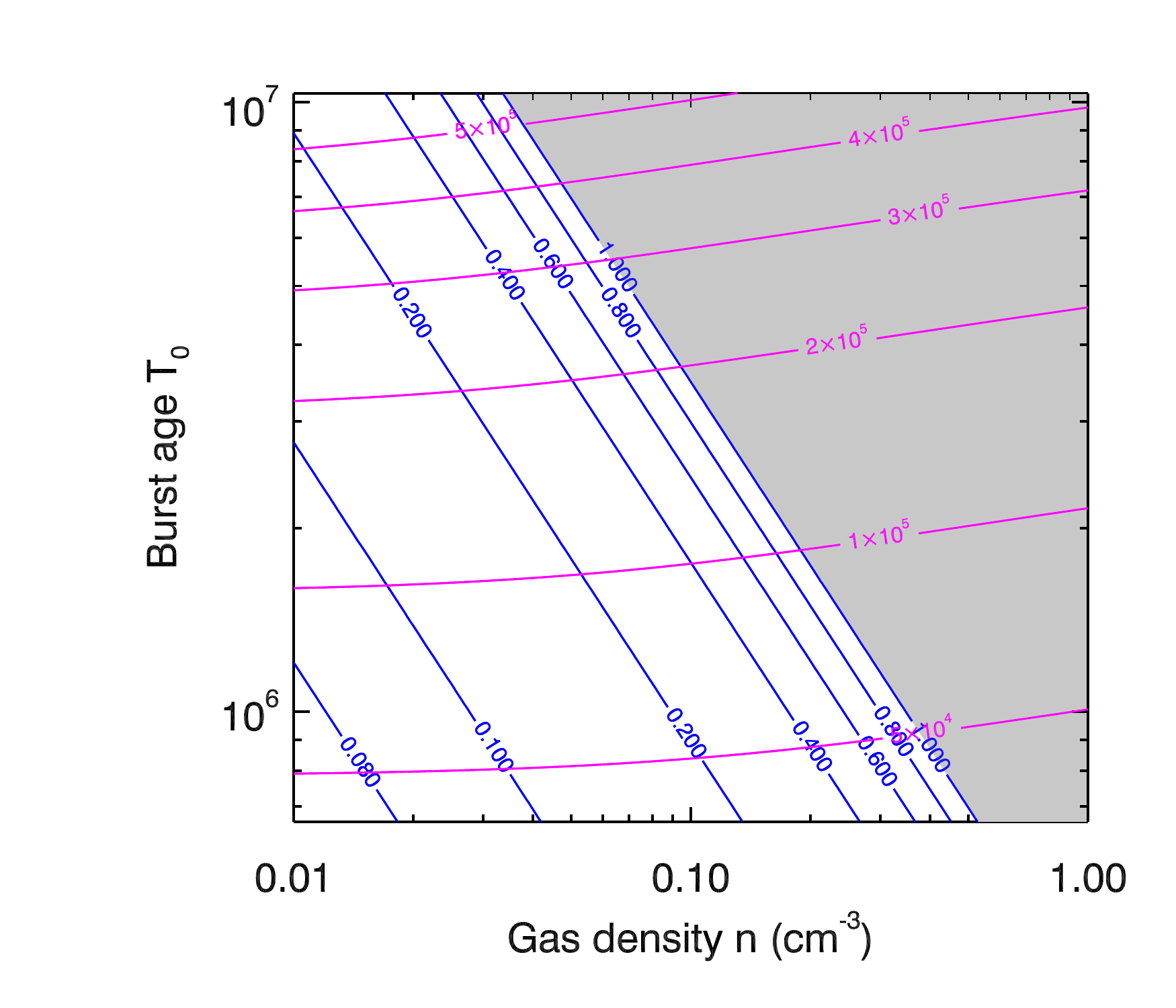}
  \caption{The constraints on the Sgr A$^\star$ burst peak Eddington
  fraction $f_{E,{\rm peak}}$ (blue contours) and burst decay
  $e$-folding time $\tau_s$ (magenta contours) as a function of Stream
  gas density $n_H$ and burst age $T_o$, for the case of very rapid
  burst decline. The grayed-out region requires $f_E >
  1$. {\it(Left):} $D=55$ kpc. {\it(Right):} $D=100$ kpc.}
   \label{f:f_edd_frac}
   \medskip
\end{figure*}

The results are shown in Fig.~\ref{f:f_edd_frac}. The grey upper
right portion of the diagrams is where $f_{E,{\rm peak}}$ exceeds 1;
this occurs sooner at higher $n_H$ (more recombination times) and larger
$T_o$ (more $e$-folding times $\tau_s$). This condition is violated
more readily for $D=100$ kpc, since a greater $\varphi_i$ is needed to
produce the same H$\alpha$ surface brightness; this also means the
minimum allowed value of $f_E$ is $\sim 3.3$ times larger. However, a
broad range of reasonable $f_E$ is allowed for burst ages greater than
$\sim 1$ Myr in both cases; a larger Stream distance favors lower $n_H$
(to increase $\tau_{\rm rec}$) and larger $f_{E,{\rm peak}}$.

Although we have assumed that $\tau_s\rightarrow 0$ in
Fig.~\ref{f:f_edd_frac}, the results are not very sensitive to this
assumption: in Appendix A, we show that $\tau_{\rm rec}/\tau_s$ is
likely to be a factor of order a few; for these values, there are modest
shifts of the curves from the instantaneous decline case (see
Fig.~\ref{f:f_edd_beta} in Appendix A). 

In our interpretation, Sgr A$^\star$ was far more active in the
past. Rapid and stochastic variations in AGN activity are to be
expected (Novak et al 2011, 2012). Depending on the Stream distance,
for plausible $n_H$ and $T_o$ the required Eddington fraction $f_E$ is
of order $0.03-0.3$ which is a factor of 10$^{7-8}$ times higher than
the quiescent state today.\footnote{While such an event would be
spectacular to behold using modern astronomical techniques, to an
ancient observer, escaping shafts of light that managed to pierce
through the heavy dust obscuration towards Earth would have been 
at least an
order of magnitude fainter than the full moon.} The most extreme event
witnessed in models by Novak et al (2011, their Fig. 6) is a transition
from $f_E \sim
10^{-3}$ to $f_E \sim 3\times 10^{-8}$ in a few Myr (45 dB). This
happens when there is a lot of material in the accretion disk around
the black hole. The AGN heats up the ISM and terminates additional
infall, and then the mass drains out of the disk with an e-folding
time of about 0.1 Myr. The timescale of the drop is roughly the
characteristic time to clear the accretion disk (G. Novak, private
communication).

But our new result demands 70-80 dB suppression within a time frame of
only $\sim 1-5$ Myr. Such a rapid variation requires an extremely
efficient and well confined `drip line' to prevent the fresh gas from
being sheared by the accretion disk which would wash out extreme
fluctuations in UV luminosity (S. Balbus, private communication).
Magnetic fields $-$ required to mediate angular momentum transport $-$
are expected to thread through a RIAF disk and
these are almost certainly needed to achieve the severe confinement
and rapid fuelling.

In time, we may learn about the detailed structure of the evolving
accretion disk before, during and after a major outburst (Ho 2008).
If the Seyfert flare model is ultimately confirmed to be the correct
explanation for the Stream's partial ionisation, it provides us with a
very interesting and spatially resolved probe of the escaping
radiation. We refrain from considering more sophisticated accretion
disk/jet models, with their attendant beaming, until more progress is
made in establishing the true source of the ionising radiation and the
Stream's trajectory. A stronger case must be made for preferring this
model over another (i.e. the shock cascade).  But we note that the
green horizontal line in Fig.~\ref{f:em} is not a good fit to most of
the data points. An inverted low amplitude parabola centred on the SGP
does better.  This can be understood in terms of an accretion disk
radiation field with a polar angle component; such models have been
presented (e.g. Madau 1988; Sim et al 2010).  But the trend to lower
Magellanic longitude conceivably can be explained if the Stream subtends a
large angle to the Galactic halo.

\subsection{UV line-driven wind} 

It is evident that the explosive nuclear activity that created the
extended X-ray, microwave and gamma-ray radiation gave rise to a
large-scale outflow from the Galactic Centre. Both starburst and AGN
activity are likely to be operating from the central regions. While
their time-averaged energetic outputs may be similar, they operate
with very different duty cycles and temporal behaviour (Alexander \&
Hickox 2012).  The Fermi observations have been discussed extensively
in the context of accretion disk activity associated with the well
established supermassive black hole (Su et al 2010; Guo \& Mathews
2012) although alternative starburst models have been presented
(Carretti et al 2013). Starbursts drive large-scale winds very
effectively and may assist with the observed bipolar activity. But as
already mentioned (\S 1; see also Appendix B), starburst activity cannot account for the
Stream H$\alpha$ emission.

Is it possible to associate the powerful radiative phase with the wind
phase?  Regrettably, there are few published accretion disk models
that provide both the ionising luminosity and mechanical luminosity of
the central source. For our discussion, we use the well prescribed
models of Proga \& Kallman (2004; hereafter P-K) that build on their
earlier work (Proga, Stone \& Kallman 2000).

In the P-K wind models, the relatively high radiation UV flux and
opacity (mainly due to line transitions) supply a strong radiative
force that is able to lift gas over the photosphere. This gas provides
significant column density to block the X-rays otherwise the wind
becomes overionised and the flow switches off.  The line-driven wind
is launched from the part of the disk where most of the UV is
emitted. In effect, the inner disk wind shields the outer wind.  The
high value of $\eta$ in eq.~(\ref{e:agn}) is consistent with our
assumption that the UV radiation dominates over X-rays and powers the
large-scale wind (Proga, Stone \& Kallman 2000).

The P-K wind reaches velocities roughly twice the escape velocity from
the launching region. So this gives velocities of about 10,000 km
s$^{-1}$ for a system with $M_\bullet = 10^8$ $M_\odot$ although maybe
somewhat less for the supermassive black hole associated with the
Galactic Centre.  With this velocity it will take only 0.1 Myr to
reach a distance of 1 kpc, and 1 Myr to reach 10 kpc. The disk wind
mass loss rate is roughly 10\% or so of the disk accretion rate and
therefore does not cause a significant reduction of the accretion
rate. The wind is unlikely to ionise cold gas at the distance of the
Stream.

Using the Proga models, Sim et al (2010) computed spectral energy
distributions as a function of viewing angle as seen from the
accretion disk.  They compute the photionisation and excitation
structure of the wind and track multiple scattering of the
photons. The polar radiation field depends on photon energy and the
escaping radiation is confined to a cone.  The X-ray and the UV
radiation come from different directions; the former propagate
parallel to the UV photosphere whereas the latter is normal to it.
Therefore the column density for the X-rays is much higher than for
the UV as expected, although some leakage is observed.

We observe that something like this may be happening in detailed
observations of nearby active galaxies. In an integral field study of
ten galactic winds, Sharp \& Bland-Hawthorn (2010) compared five
starbursts and five AGNs.  The AGN winds show clear evidence for
non-thermal ionisation from the central source across the wind
filaments to the radial limits of the data.  But AGN ionisation cones
are not always associated with winds.  For example, the most famous of
the Seyfert ionisation cones is NGC 5252 (Tadhunter \& Tsvetanov 1989)
which is not associated with an energetic outflow.  In the context of
the P-K model, we associate these cones with AGNs where strong X-rays
escape from the nucleus which serve to suppress the line-driven
wind. This distinction may become less clear cut if more powerful
line-driven winds (presumably from more massive black holes) are able
to drive shocks in the gas along the ionisation cone. Thus ionisation
cones may be detectable in X-rays even while the central source
irradiates the cone exclusively with UV. Shocked gas tends to radiate
at a higher temperature compared to photoionised gas, and this may
allow these cases to be separated.

Line-driven winds struggle with black hole masses as low as that
associated with Sgr A$^\star$ unless the accretion rate is close to
the Eddington limit.  If the Stream ionisation is due to a burst of
radiation from a P-K disk, then $f_E \sim 1$ is an order of magnitude
more than is need to account for the observed H$\alpha$ emission for
the canonical Stream distance (55 kpc), although it would aid
ionisation of the Stream at the larger distance.  In principle,
$f_{\bullet,{\rm esc}}$ could be lower than our assumed value of
100\%. But the high limit is consistent with what we know about
ionisation cones (e.g. Mulchaey et al 1996) and is a consequence of
the P-K wind model where the wind has cleared a channel for the UV
emission (Proga \& Kallman 2004; Sim et al 2010).

\section{Conclusions}

We have shown how the Magellanic Stream is lit up in optical emission lines
at a level that cannot be explained by disk or halo sources.  
A possible explanation is a shock cascade caused by the break-up of
clouds and internal collisions along the Stream (Bland-Hawthorn et al
2007) but this becomes untenable if the Stream is much further than
the canonical distance of 55 kpc. 

We have introduced time-dependent ionisation calculations with {\it MAPPINGS IV}
for the first time in order to present a promising Seyfert flare model 
that adequately explains the observed photoionisation levels along
the Magellanic Stream. The model works at both the near and far Stream
distances, and can be tested in future observations.  A `slow shock
cascade' is expected to produce a steeper Balmer decrement
(H$\alpha$/H$\beta$ $>$ 3.1) than the flare model. Since the
Magellanic Stream has a very low dust fraction ([Fe/H]$\approx$-1),
this is likely to be the most accessible discriminant between the
models. Other useful diagnostics (\HeII/H$\beta$, \OIII/H$\beta$)
reach peak values shortly after the Seyfert flash but fade rapidly.

We cannot yet identify the specific event which triggered the burst
of Seyfert activity although the stellar record tells us the past 10 Myr
have been very active (Ponti et al 2013). The time lag between an
accretion event and the onset of starburst or AGN activity (or how
these operate together) is a major unsolved problem in astrophysics.
The inner tens of parsecs provide 
many possible cloud candidates, assuming it was not largely
consumed, many on highly elliptic orbits. If our model is correct,
it provides many new challenges for 
the burgeoning field of Galactic Centre research.
Regardless of the origin of the emission, the Stream provides an
important constraint on past AGN activity and on models that attempt
to explain the {\it Fermi} gamma-ray bubbles.

\section{Acknowledgment}

This work came about during the April 2013 ``Fermi Bubbles'' workshop
held at KIPAC Stanford organised by Dmitry Malyshev and Anna
Frankowiac. We acknowledge insightful conversations with Steve Balbus,
Bill Mathews, Daniel Proga, Greg Novak, Bruce Draine and Kyler Kuehn. JBH is
indebted to James Binney for early comments on the manuscript. We are
particularly grateful to Roger Blandford and the participants of the
Kavli workshop without whom this paper may not have been realised.
We thank the referee for suggestions that led to more detailed discussions
in the appendix.

\newpage
\appendix
\section{A. A simple model for time-dependent evolution of the ionisation
  fraction and H$\alpha$ surface brightness}

Consider the following simple model for the Stream clouds: a uniform
density gas of pure hydrogen, with a photon flux $\varphi_i$ normally
incident upon it. If the gas has been exposed to the ionising photons
for long enough to reach ionisation equilibrium, then all of the
photons will be absorbed in a length $L$ given by
\bee
\alpha n_e^2 L = \varphi_e
\eee
where $\alpha$ is the recombination coefficient. This is just the condition
that the column recombination rate equals the incident flux. Thus
\bee
L = {\varphi_i\over \alpha n_e^2}
\eee
To simplify even further, assume that for depth $d < L$ into the
Stream gas, the gas is completely ionised, while for $d > L$, it is
neutral. Hence all the emission measure comes from $ d < L$, and we
will also ignore any effects of absorption on $\varphi_i$, so the region
with $d < L$ can be treated as uniform.

Now suppose that the ionisation rate decreases from the initial value
for which the equilibrium was established. Without loss of generality,
we can assume an exponential decline for $\varphi_i$, with a characteristic
timescale for the ionising source $\tau_s$ (Sharp \& Bland-Hawthorn 2010). 
The time-dependent equation for the electron fraction $x_e = n_e/n_H$ is
\bea
{dx_e\over dt} &=& -\alpha n_H x_e^2 -\zeta x_e + \zeta \cr
&=& -\alpha n_H x_e^2 + \zeta_0 e^{-t/\tau_s}(1-x_e)
\label{e:dxdt}
\eea
where $\zeta$ is the ionisation rate per atom. 

Consider first the case where $\tau_s \rightarrow 0$, so that $\varphi_i$
declines instantaneously to zero. Then the second and third terms in
eq.~(\ref{e:dxdt}) vanish, and we just have
\bee
{dx_e\over dt} = -\alpha n_H x_e^2
\eee
This is easily solved with the substitution $u = x_e^{-1}$, and with
the initial condition $x_e=1$ at $t=0$ we get
\bee
x_e = {1\over 1 + \alpha n_H t}
\eee
Defining the recombination timescale 
\bee
\tau_{\rm rec} = 1/\alpha n_H
\eee
this is simply
\bee
x_e = \left(1+t/\tau_{\rm rec}\right)^{-1}
\label{e:ionfrac}
\eee
To evaluate eq.~(\ref{e:ionfrac}) for the conditions in the Stream, we use
$\alpha_B = 2.6\times 10^{-13}$ 
cm$^3$ s$^{-1}$ for the recombination coefficient (appropriate for
hydrogen at $10^4$ K), and 
use the fiducial values $\varphi_i = 10^6\varphi_6$ phot s$^{-1}$, $n_H= 0.1
n_{-1}$ cm$^{-3}$. Then
\bee 
L= 125 {\varphi_6\over n_{-1}^2} \; {\rm pc}
\eee
\bee 
\tau_{\rm rec} = 1.2\times 10^6/n_{-1}\; {\rm yr}
\label{e:Trec}
\eee
and the emission measure 
\bee
{\cal E}_m = L n^2 x_e^2 = 1.25 \varphi_6 x_e^2(t)\; {\rm cm^{-6}\;pc}
\eee
so the gas density enters explicitly only through the recombination
time. The resulting H$\alpha$ emission will be
\bee
\mu_{\rm H\alpha} = 413 \varphi_6 x_e^2(t)\; {\rm mR}
\eee
or, with eq.~(\ref{e:ionfrac})
\bee
\mu_{\rm H\alpha} = 413 \varphi_6 (1+t/\tau_{\rm rec})^{-2}\; {\rm mR}
\eee

\begin{figure}[htbp]
  \includegraphics[scale=0.55,clip=true,trim=0mm 21.5mm 0mm 12.5mm]{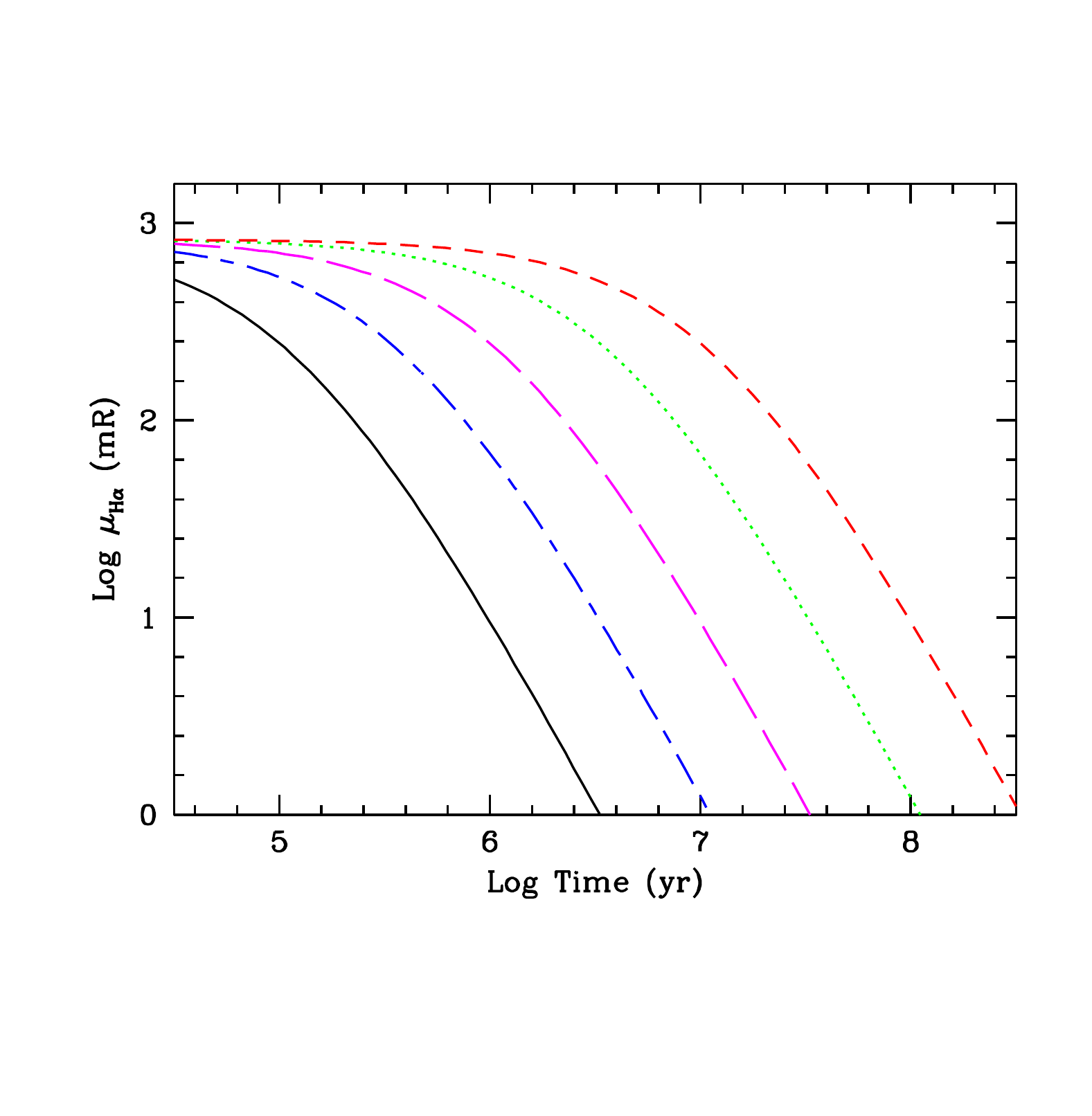}
  \includegraphics[scale=0.55,clip=true,trim=0mm 0mm 0mm 12.5mm]{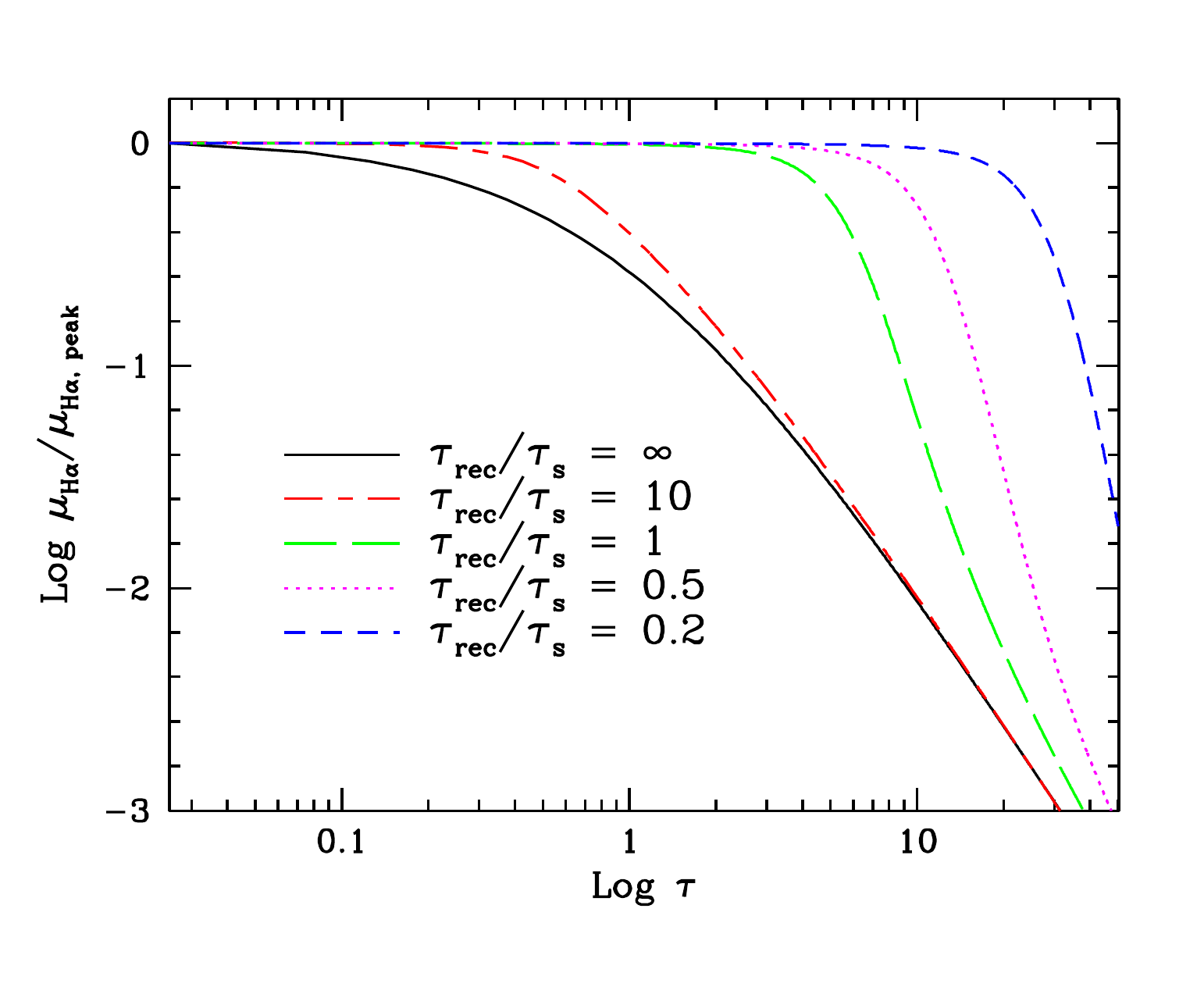}
  \caption{{\it (Left)} The H$\alpha$ surface brightness as a function
  of time predicted by eq.~(A12), for $\varphi_6=2$. From left to
  right, the curves are for gas density $n_H=1$, 0.3, 0.1, 0.03, and
  0.01 cm$^{-3}$. {\it (Right)} The evolution of the H$\alpha$ surface
  brightness (scaled to the peak brightness)
  with dimensionless time $\tau$ obtained by solving
  eq.~(A16) for several values of the ratio $\beta$ of
  recombination time to ionising photon flux decay time. Curves are
  labeled with $\beta$; all models assume a ratio of recombination 
  time to
  ``$t=0$ ionisation time,'' $\gamma = \tau_{\rm rec}/\tau_i^0 = 240$. }
   \label{f:halpha_evolve}
   \medskip
\end{figure}

In the lefthand panel of Fig.~\ref{f:halpha_evolve}, we plot the
prediction of eq.~(A12) for the H$\alpha$ surface brightness as a
function of time for $\varphi_6=2$, as used in eqs. (10) and (11)
for $D=55$ kpc. Comparison with the left-hand panel of Fig.
\ref{f:map_Em}\ shows that this simple model agrees well with the
detailed MAPPINGS results, except for the highest densities. The
discrepancy is largely because eq.~(A12) predicts that $x_e$
depends only on $t/\tau_{\rm rec}$, and hence $x_e$ remains close to
unity (and thus $\mu_{{\rm H}\alpha}$ at its peak value) only if
$t/\tau_{\rm rec}$ is small, but that is not true for the highest
densities at the earliest times for the range of times that are
plotted. However, it clearly does a good job of reproducing the
late-time behaviour ($\mu_{{\rm H}\alpha}\propto (t/\tau_{\rm
rec})^{-2}$), to which all the models in Fig.~\ref{f:map_Em}
asymptote.

Equation (A3) does not have an analytic solution when the
time-dependence of $\varphi_i$ is included. However, it is easily solved
numerically and can be transformed into a more useful form with some
trivial definitions. Define the dimensionless time $\tau$ by 
\bee
\tau\equiv \alpha n_H t = t/\tau_{\rm rec}\;;
\eee
$\tau$ is simply the time measured in units of the recombination
time. In addition, define
\bee
\gamma = \tau_{\rm rec}/\tau_i^0
\eee
where $\tau_i^0$ is the ionisation time at $t=0$, and
\bee
\beta = \tau_{\rm rec}/\tau_s\;,
\eee
the ratio of recombination to ionising photon luminosity $e-$folding times.
Then eq.~(\ref{e:dxdt}) becomes
\bee
{dx_e\over d\tau} = -x_e^2 + \gamma e^{-\beta\tau}(1-x_e)
\eee
We can write the ionisation rate per H atom
as
\bee
\zeta \simeq C_i \varphi_i\sigma_0
\eee
where $\sigma_0$ is the H ionisation cross-section at threshold and
$C_i$ is a constant of order unity that depends on the shape of the
spectrum. The ionisation time $\tau_i^0$ then evaluates to
\bee
\tau_i^0 = 5000{C_i\over\varphi_6}\;{\rm yr}
\eee
and we can write $\gamma$ as (using eq.~[A9] for
$\tau_{\rm rec}$)
\bee
\gamma = 240{\varphi_6\over C_i n_{-1}}
\eee

The resulting H$\alpha$ surface brightness obtained from the solution
of eq.~(A16) for the ionisation fraction, normalized to the peak
value, is shown in the righthand panel of Fig.~\ref{f:halpha_evolve}
for $\gamma=240$ and values of $\beta$ from 0.2 to $\infty$ ($\tau_s
\rightarrow 0$, the case shown in the lefthand panel). However, an
important point from this analysis can be derived simply from the form
of eq.~(A16). As just shown, $\gamma$ must be large -- this is inevitable
from the assumption that the gas in the H$\alpha$-emitting region is
highly ionised to begin with. Hence the ionisation fraction (and thus
the H$\alpha$ surface brightness) will not begin to decrease
substantially until
\bee 
e^{-\beta\tau} \sim 1/\gamma 
\eee 
and thus until $\tau$ reaches the critical value 
\bee 
\tau_c \sim {\ln\gamma\over\beta} .
\eee
Physically, this is just a reflection of the requirement that the
ionisation time must be longer than the recombination time before the
ionisation fraction begins to drop. If $\beta\lesssim 1$ --- the
ionising photon flux is decreasing on a timescale longer than the
recombination timescale --- the ionisation fraction (and thus the
H$\alpha$ emission) will not begin to decline substantially until many
recombination times have passed.

The numerical solutions of eq.~(A16) show that the expression (A21)
for the critical time is quite accurate: for $\gamma=240$, it predicts
$\tau_c \sim 0.55$, 5.5, 11, and 27 for $\beta = 10$, 1, 0.5, and 0.2,
respectively.  Since $\tau_c$ depends only logarithmically on $\varphi_i$
and the gas density $n_H$, the precise values of these quantities are
unimportant --- all that matters is that, generically, $\ln\gamma
\sim$ a few.  Unless $\tau_s$ is much shorter than $\tau_{\rm rec}$
($\beta \gg 1$), the decline of $x_e$ --- and thus of $\mu_{{\rm
H}\alpha}$ --- is substantially delayed from the instantaneous
$\varphi_i$ turn-off case. (Note also that for $\beta < 1$, the
  decline is steeper once it begins. This is because the derivative of
  $x_e$ with respect to $\ln\tau$ has its maximum at $\tau_c$ ---
  physically, there is simply more time available between the steps of
  $\ln\tau$ at these later times.)

This simple model can also be used to address another very important
issue. We do not know {\it a priori} what the peak luminosity of the
burst was, or the timescale on which it decayed. All we know is that
the peak H$\alpha$ surface brightness was at least equal to the
present-epoch value. Define
\bee
\rho = \mu_{{\rm H}\alpha, \rm obs}/\mu_{{\rm H}\alpha, \rm peak}
\eee
which is also equal to the ratio of minimum to peak ionizing photon
luminosity $N_{i,{\rm min}}/N_{i,{\rm peak}}$ and to the ratio of the
minimum required Eddington fraction to the peak value, $f_{E,{\rm
    min}}/f_{E,{\rm peak}}$. Consider first the limit of
$\tau_s\rightarrow 0$. From eq.~(A7), we have
\bee
\rho = \left(1+t/\tau_{\rm rec}\right)^{-2}
\eee
(cf. eq.~[A12]). The time needed for the H$\alpha$ surface
brightness to decline to its observed value, in units of the
recombination time, is simply
\bee
\tau_\rho = \rho^{-1/2} - 1
\eee
The value of $\tau_\rho$ predicted by eq.~(A24) agrees reasonably
well with the models presented in Fig.~\ref{f:map_Em}. For $D=55$
kpc, $\rho\simeq 0.2$, and so $\tau_\rho=1.24$. Using eq.~(A9) for
$\tau_{\rm rec}$, the values from Table 1 give $\tau_\rho=1.08-1.2$, while
for the $D=100$ kpc model, $\rho\simeq 0.64$, so the predicted value
of $\tau_\rho = 0.25$, while the derived values range from 0.25 to
0.35. The differences between the prediction and the calculated values
result from the simplification of eq.~(A9) in assuming a constant
recombination coefficient that is independent of time and ignores the
different temperature histories as shown in Fig.~\ref{f:map_T}.

For the case of non-instantaneous decline of the burst luminosity, we
can easily solve for $\tau_\rho$ numerically for different values of
$\beta$. One difference from the results shown in the righthand panel
of Fig.~\ref{f:halpha_evolve} is that we must define $\gamma$
consistently with the choice of $\rho$; this can be seen by noting
that eq.~(A14) for $\gamma$ can be written, using eq.~(2)
for $\varphi_{i,{\rm min}}$ and eq.~(A9) for $\tau_{\rm rec}$, as 
\bee
\gamma = {93.6\over \rho n_{-1}}
\eee
which we use to specify $\gamma$ as a function of $\rho$.

\begin{figure}[htbp]
  \includegraphics[scale=0.55,clip=true,trim=0mm 0mm 0mm 0mm]{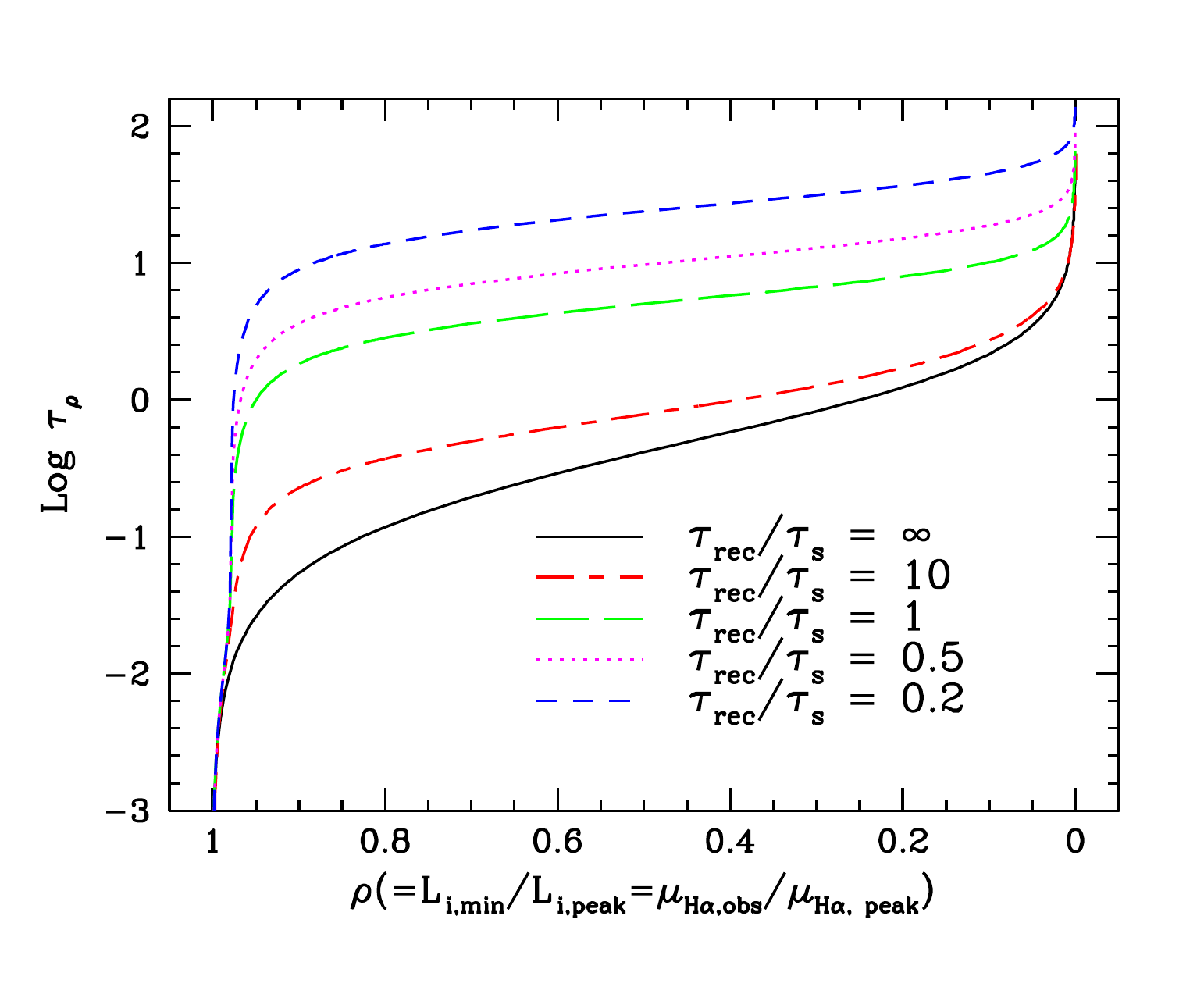}
  \includegraphics[scale=0.55,clip=true,trim=0mm 0mm 0mm 0mm]{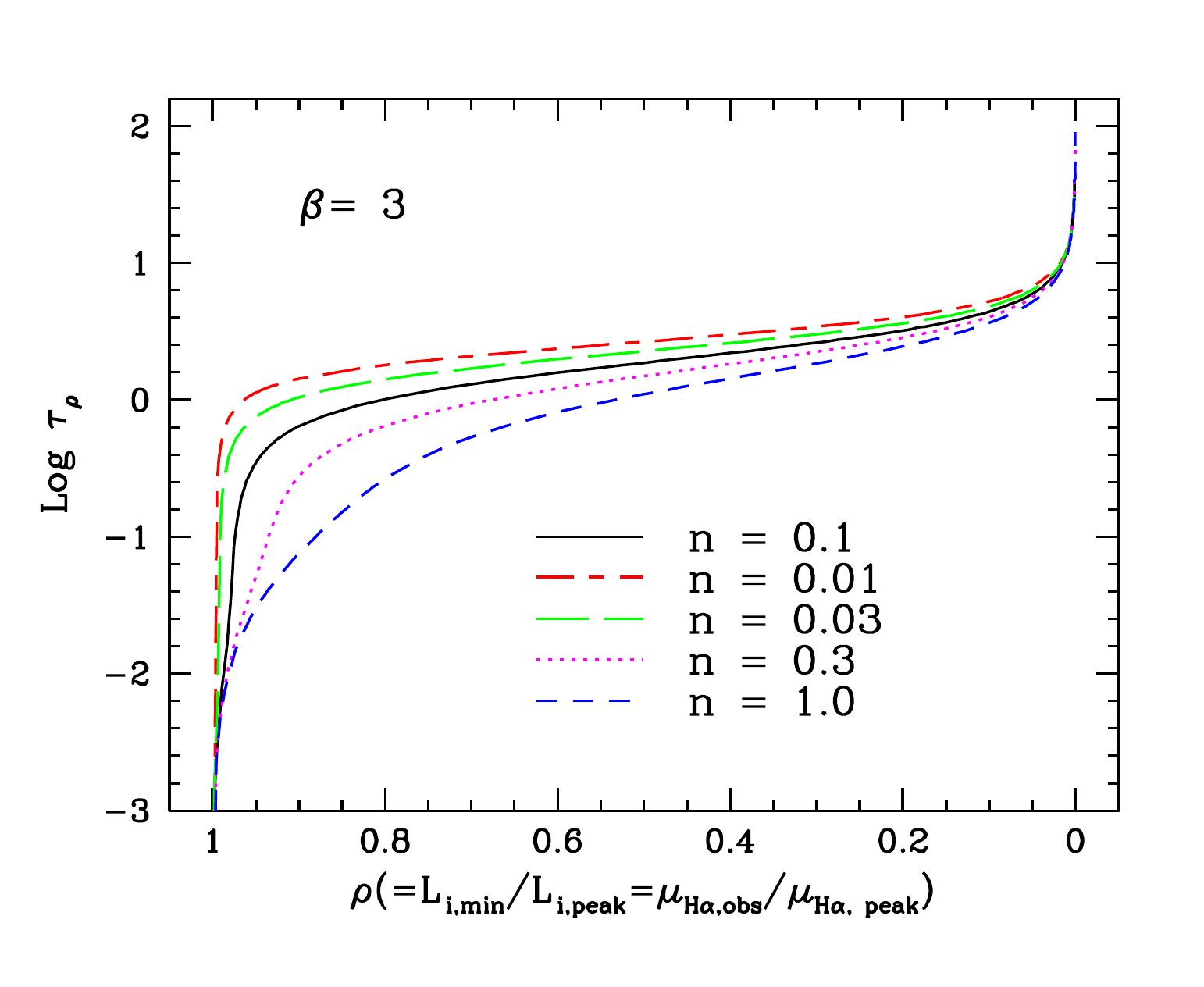}
  \caption{{\it (Left)} The dimensionless time $\tau$ required for the
  H$\alpha$ surface brightness to decline to a fraction $\rho$ of its
  peak value; $\rho$ is also equal to the ratio of the minimum
  ionizing photon flux or luminosity to their respective peak values.
  Curves are labeled with $\beta$, the ratio of recombination time to
  ionising photon flux decay time. A Stream gas density of $n_H=0.1$
  cm$^{-3}$ was assumed. {\it (Right)} As in the left panel, except
  for $\beta$ fixed at 3 and different values of the gas density (labeled).}
   \label{f:tau_rho}
   \medskip
\end{figure}

The solutions for $\tau_\rho$ are shown in Fig.~\ref{f:tau_rho}. The
left panel assumes a Stream gas density fixed at $n_H=0.1$ cm$^{-3}$, and
shows the results for several different values of $\beta$ (as in
Fig. 7). The offset between the curves with different $\beta$ is a
direct reflection of the delay in the decline of $\mu_{{\rm H}\alpha}$
seen in the righthand panel of Fig.~\ref{f:halpha_evolve}. In the
righthand panel of Fig.~\ref{f:tau_rho}, $\beta$ has been fixed at 3
(see below), and $\tau_\rho$ is plotted against $\rho$ for several
different densities. From eq.~(A25), for fixed $\rho$ the value
of $\gamma$ increases with decreasing density $n_H$, which is why the
curves flatten out as $n_H$ declines to the lowest values. The spread is
much smaller than in the variable-$\beta$ curves shown in the lefthand
panel, especially for small values of $\rho$; this is because $\tau_c$
depends only on $\ln\gamma$ whereas it depends linearly on $1/\beta$,
as discussed above (eq.~[A21]). The steep decline as $\rho\rightarrow 1$ seen in both panels of Fig.~10 is imposed by the initial condition $x_e=1$ at $\rho=1$. The convergence of all the models to the same steep rise as $\rho\rightarrow 0$ results from the negligible size of the ionization term at late times ($\tau\gg\tau_c$), so that they all approach the $x_e\propto \tau^{-2}$ solution (A7) for the instantaneous-decline case.

\begin{figure}[htbp]
  \includegraphics[scale=0.55,clip=true,trim=10mm 0mm 0mm 5mm]{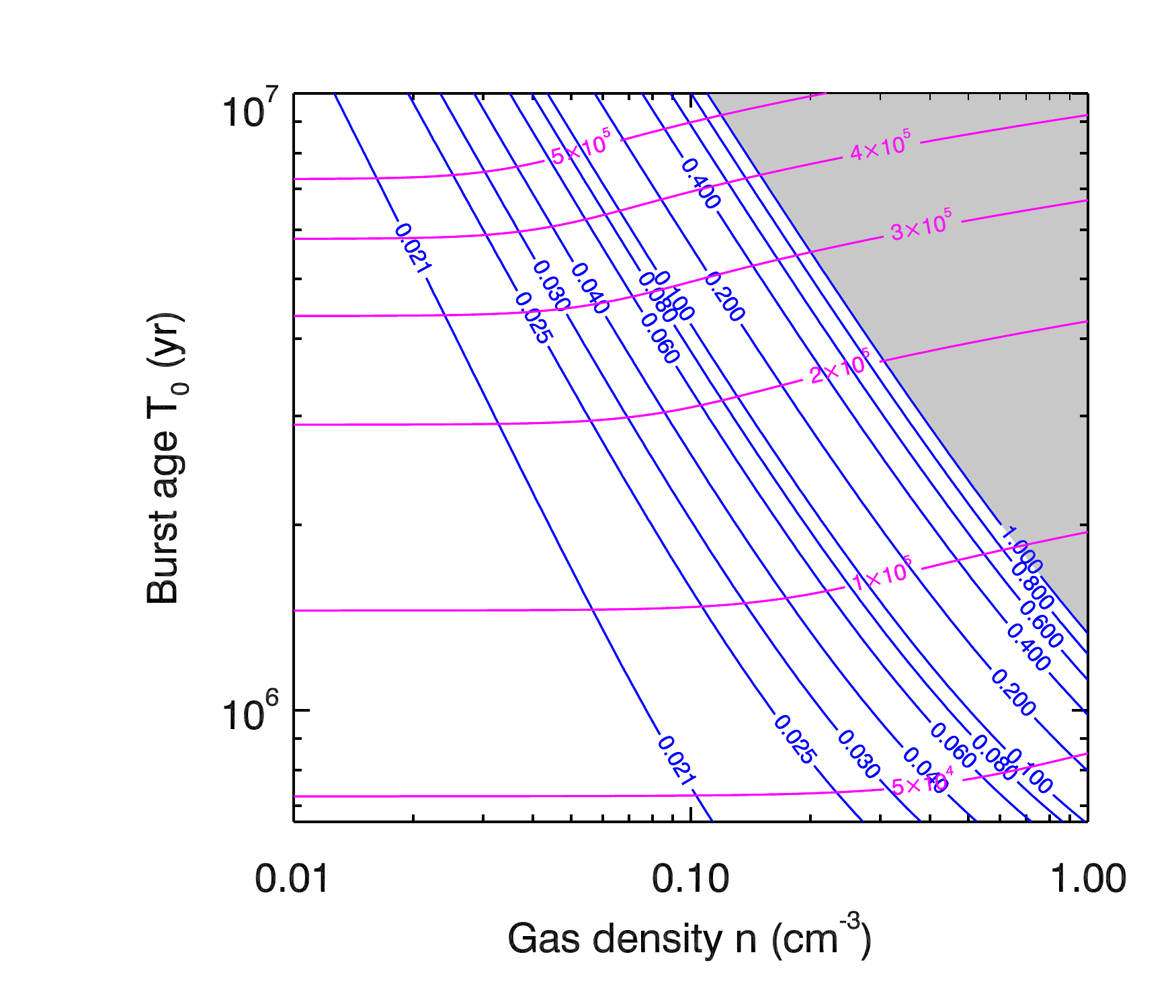}
  \includegraphics[scale=0.55,clip=true,trim=20mm 0mm 0mm 5mm]{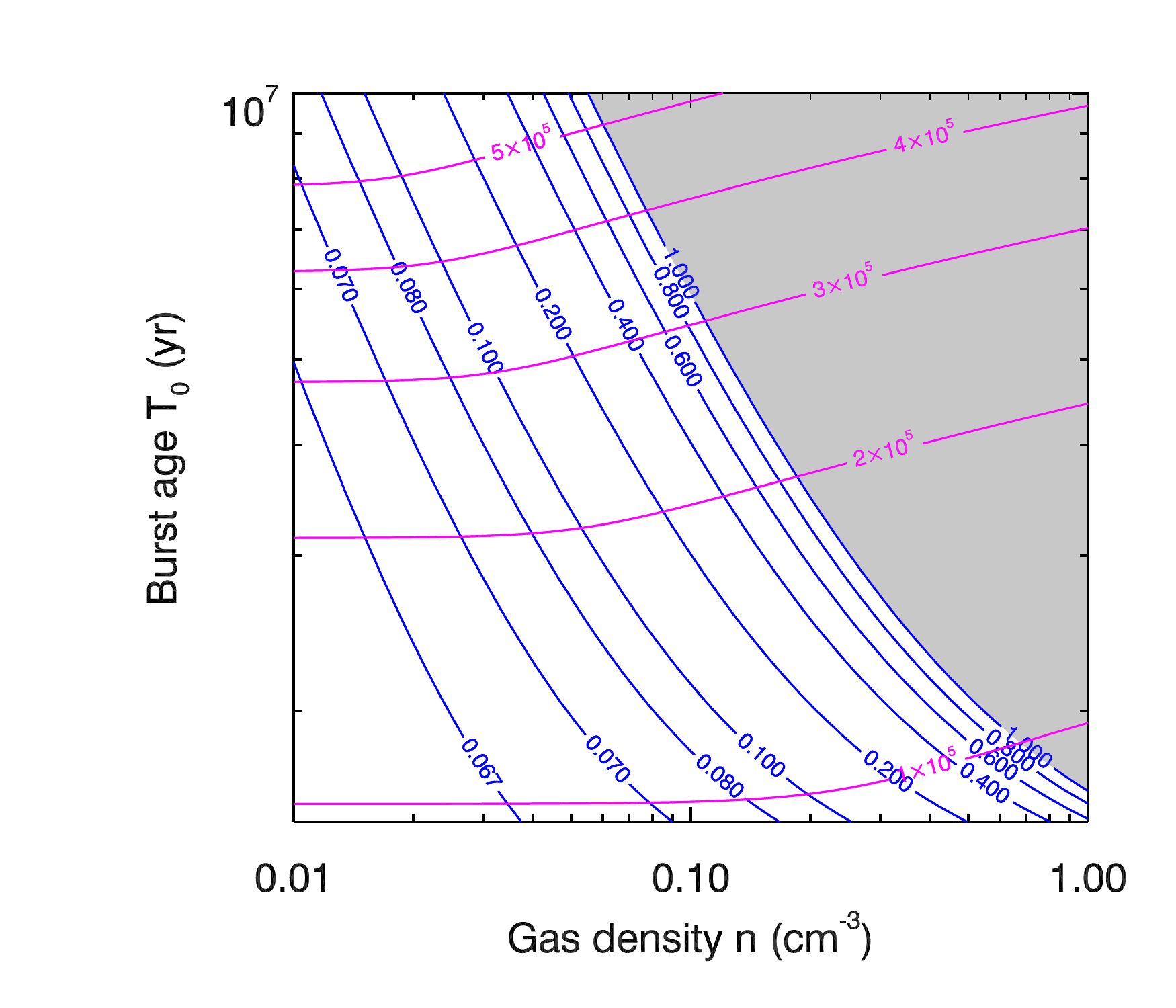}
  \caption{The required Eddington fraction as a function
  of burst age $T_o$ and Stream gas density $n_H$. {\it (Left)} For
  $D=55$ kpc and $\beta =3$. {\it (Right)} For $D=100$ kpc and
  $\beta=6$. Compare with Fig.~\ref{f:f_edd_frac}.}
   \label{f:f_edd_beta}
   \medskip
\end{figure}

We use these results in \S 5.1 to discuss the constraints on the peak
luminosity and decay timescale of the Sgr A$^\star$ flare. Here we note
that $\beta$ is likely to be at least a few. In terms of the Eddington
fraction and the burst age $T_o$, we can evaluate eq.~(A24) to get
\bee
n_H T_o \lta 1.2\times 10^5\left({7.1\over (D/55\,{\rm
    kpc})}-1\right) \ \ \ {\rm cm}^{-3} \; {\rm yr}
\eee
for $f_{E,\rm peak}=1$. This gives the largest possible value for
$\tau_s$. In \S 5.1, we infer a central flare that has faded by 80 dB,
or approximately 18 $e$-folding times, since the burst peak. With eq.~(A26), 
we then get that $\tau_s \lta 4(2)\times 10^5/n_H$ for
$D=55(100)$ kpc. This implies $\beta \gta 3 - 6$ for the Stream
distances: burst decay times longer than $\tau_s \sim$ a few $\times
10^5$ yr are unlikely, given the probable age of the {\it Fermi}
bubbles. In Fig.~\ref{f:f_edd_beta} we show the required value of 
Eddington fraction $f_E$ as a function of burst age $T_o$ and gas
density $n_H$ for these two cases. Comparison with Fig.~\ref{f:f_edd_frac}
in \S 5.1 shows that the differences from the case
$\beta\rightarrow\infty$ are modest.

We mention briefly one further important point. The gas
recombination/cooling times can obscure any natural variations in the
source ionising luminosity. As discussed in the text, the Stream
H$\alpha$ emission {\it must} arise from a fading source, but the
fading time of the H$\alpha$ emission is limited by the gas
recombination time, which imposes a transfer function on the
luminosity variations. The source variation timescale $\tau_s$ could
in principle be shorter, and the luminosity variations even more
dramatic, than what we infer. In reality, the transfer function will
be even more complex than what is implied by the simple model used
here, since the real Stream gas has distributions of gas density and
column density. In general, when comparing numerical models of AGN
variability with the Stream emission (e.g. Novak et al 2011), the
modelled data stream must be temporally convolved with a function
whose bandwidth will depend on the gas density.

\section{B. The implausibility of a starburst origin of the flare}

Equation (3) provides a minimum estimate for the ionising photon
luminosity needed to explain the Stream H$\alpha$ emission of
$N_i\sim$ a few $\times 10^{53}$ phot s$^{-1}$ --- this assumes no
fading of the emission and no significant absorption of the ionising
photon flux. What starburst parameters does this imply?

Maloney (1999) quotes a ratio of ionising photon luminosity to star
formation rate of
\bee
N_i/\dot M_* \sim 10^{53} \;{\rm phot\;s^{-1}}\;M_\odot^{-1}\;{\rm yr}
\eee
with substantial caveats on starburst age, upper and lower mass
cutoffs, etc. This number agrees very well with the properties of the
massive young star clusters formed in the Galactic Centre over the
past several Myr: the Quintuplet, the Arches, and the Nuclear Cluster,
which have total stellar masses $\sim 10^4M_\odot$, ages in the range
1--7 Myr, and $N_i\sim 10^{51}$ phot s$^{-1}$ (Figer,
McLean, \& Morris 1999). Even for burst timescales as short as 1 Myr,
the resulting star formation rates $\dot M_* \sim 0.01$ $M_\odot$
yr$^{-1}$, giving ionising photon luminosities per unit SFR in
agreement with (B1). We can also use these observations to estimate
the ionising photon flux per unit mass of stars formed: this is
\be
N_i/M_*\sim 10^{47} \;{\rm phot\; s^{-1}}\; M_\odot^{-1}\;.
\ee
Powering the Stream emission at the minimal levels of eq.~(3)
thus requires a SFR of
\bee
\dot M_* \sim 1.4 - 4.7\;M_\odot\;{\rm yr^{-1}}
\eee
and a total mass of stars formed of 
\bee
M_* \sim (1.4 - 4.7)\times 10^6 M_\odot
\eee
The requirement of eq.~(B3) exceeds by $\sim$ two orders of
magnitude all estimates of the SFR in the Galactic Centre within the
last $1-100$ Myr (e.g. Pfuhl et al 2011, their Fig. 14, and many
references therein; see also above). A similar problem arises with the
mass of stars in the Galactic Centre as a function of age (Pfuhl et al 2011). This
number for the SFR is likely to be a substantial underestimate, since
we have neglected extinction in the vicinity of the star-forming
regions: for the three young Galactic Centre clusters discussed above, a
large fraction of the emitted ionising photons are absorbed locally.

In fact, the situation is even worse than this: any such nuclear
starburst would have to have declined in luminosity by $\sim 2$ orders
of magnitude from the required peak luminosity to the present epoch,
indicating that $\sim 5$ or more $e$-folding times have elapsed. For
plausible minimum starburst timescales ($\tau_s \sim 2-3$ Myr), this
makes the burst epoch too early to match the age of the {\it Fermi}
bubbles. Except for implausibly small Stream densities ($n_H\sim 0.01$
cm$^{-3}$), this also indicates that $\beta \lesssim 1$, and even
though this will delay the decline of H$\alpha$ surface brightness
compared to the case where the flare shuts off in a time $\tau_s \ll
\tau_{\rm rec}$ (see Appendix A), it also introduces a fine-tuning
problem: unless we are catching the Stream emission at a time very
close to $\tau_c$ as given by eq.~(A21), the observed $\mu_{{\rm
H}\alpha}$ will be substantially less than the peak value, indicating
that the peak SFR and the mass of stars formed in the burst would need
to be even larger than the estimates of eqs. (B3) and (B4). Hence
starburst models for the Stream H$\alpha$ emission are simply not
viable: {\it the required star formation rates greatly exceed anything seen
in the star formation history of the Galactic Centre.}

\section{C. Spectral energy distribution of an accretion disk}

Our model for the accretion disk comprises a `cool' big blue bump 
and a `hot' power law component.  We define the specific photon luminosity 
for the two-component spectrum by
\bee
N_\bullet = k_1 (E/E_1)^{-2/3} e^{-E/E_1} + k_2 (E/E_2)^{-\alpha}
e^{-E/E_2} {\cal H}(E-E_1) {\quad \rm phot\ s^{-1}\ eV^{-1}}\,, 
\eee
where ${\cal H}[E-E_1] = 1$ if $E>E_1$ and ${\cal H}[E-E_1] = 0$ otherwise.
Then the {\it total} AGN luminosity is given by
\bea
L_\bullet &=& \int_0^\infty E N_\bullet dE \cr
&=& k_1 E_1^2 \int_0^\infty \epsilon^{1/3} e^{-\epsilon} d\epsilon
+ k_2 E_2^2 \int_{w_2}^\infty \mu^{1-\alpha} e^{-\mu} d\mu \cr
&=& L_0 + L_2 
\eea
where $\epsilon\equiv E/E_1$, $\mu\equiv E/E_2$, and $w_2\equiv E_1/E_2$.

Taking the hydrogen ionisation potential, $I_H = 13.59844$~eV, and
$w_1 = I_H/E_1$, the AGN ionising luminosity is found by integrating
from the Lyman limit to infinity, 
\bea
L_{\bullet,i} &=& \int_{w_1}^\infty E N_\bullet dE \cr
&=& k_1 E_1^2 \int_{w_1}^\infty \epsilon^{1/3} e^{-\epsilon} d\epsilon
+ k_2 E_2^2 \int_{w_2}^\infty \mu^{1-\alpha} e^{-\mu} d\mu \cr
&=& L_1 + L_2 
\eea
where the limit for the second integral remains the same since $w_2 > w_1$.

The big blue bump {\it total} contribution is
\bee
L_0 = k_1 E_1^2 \; \Gamma\left({4\over 3}\right)
\label{e:l0}
\eee 
where $\Gamma(a)$ is the complete gamma function.
The big blue bump ionising contribution is
\bee
L_1 = k_1 E_1^2 \; \Gamma\left({4\over 3}, w_1\right)
\label{e:l1}
\eee
and, finally, the third integral (the power-law X-ray $+$ gamma-ray
contribution) is 
\bee 
L_2 = k_2 E_2^2 \; \Gamma(2-\alpha, w_2)
\label{e:l2}
\eee
where we use the incomplete gamma function, $\Gamma(a, b)$, and
$\alpha$ must be less than 2. Here we adopt a photon spectral index of
$\alpha=1.9$. 

If we define $\eta\equiv L_1/L_2$, so that $L_2 = L_{\bullet,i}
/(1+\eta)$ and $L_1 = \eta L_2$, then we can write $k_1$ and $k_2$ as
\bea
k_1 &=& L_{\bullet,i} E_1^{-2} \left[\frac{\eta }{\Gamma(4/3,
    w_1)(1+\eta)} \right]  \, , \\ 
\label{e:k1}
k_2  &=& L_{\bullet,i} E_2^{-2} \left[ \frac{1}{\Gamma(2-\alpha, w_2)
    (1+\eta) } \right] \, . 
\label{e:k2}
\eea
The scaling coefficient ratio $k_1/k_2$ is independent of
$L_{\bullet,i}$ such that 
\bee
{k_1\over k_2}  = \eta \left[ \frac{E_2^2}{E_1^2} \right]
\frac{\Gamma(2-\alpha, w_2)}{\Gamma({4/3}, w_1)} \, . 
\label{e:kratio}
\eee
Once the AGN luminosity $L_\bullet$ and UV to X-$\gamma$ ratio
($\eta$) are specified, the normalisation constants $k_1$ and $k_2$
follow immediately (see \S 2).

\end{document}